\documentclass{aa}
\usepackage{color}
\usepackage{amssymb}
\usepackage{amsmath}
\usepackage{graphicx}
\usepackage{txfonts}
\usepackage{natbib}
\usepackage{lscape}

\begin{document}

\title{The relative and absolute timing accuracy of the EPIC-pn camera on \emph{XMM-Newton}, from X-ray pulsations of the Crab and other pulsars}

\author{A. Martin-Carrillo \inst{1,2}, M.G.F.\ Kirsch\inst{3}, I.\ Caballero\inst{4}, M.J.\ Freyberg\inst{6}, A. Ibarra \inst{2}, E.\ Kendziorra\inst{5}, U. Lammers\inst{2} , K. Mukerjee\inst{7}, G.\ Sch\"onherr\inst{8}, M.\ Stuhlinger\inst{2}, R.D.\ Saxton\inst{2},
R.\ Staubert\inst{5}, S. Suchy\inst{9},
  A.\ Wellbrock\inst{10}, N. Webb\inst{11,12}, M. Guainazzi\inst{2}}

\offprints{\\ A. Martin-Carrillo, antonio.martin-carrillo@ucd.ie \\
M.G.F.\ Kirsch, Marcus.Kirsch@esa.int}

\institute{ University College Dublin, School of Physics, Space Science Group, Belfield, Dublin 4, Ireland
\and
European Space Astronomy Centre (ESAC), ESA, 
P.O. Box 78, 28691 Villanueva de la Ca\~nada, Madrid, Spain
\and
European Space Operations Centre (ESOC), ESA, 
Robert-Bosch-Str. 5 D - 64293, Darmstadt, Germany
\and
 CEA Saclay, DSM/IRFU/SAp -- UMR AIM (7158) CNRS/CEA/Universite P.Diderot  F-91191 Gif sur Yvette France 
 \and
  Institut f\"ur Astronomie und Astrophysik der Universit\"at T\"ubingen,
  Abteilung Astronomie, Sand 1, 72076 T\"ubingen, Germany
\and
  Max-Planck-Institut f\"ur extraterrestrische Physik,
  Giessenbachstrasse, 85748 Garching, Germany
  \and
  Department of Astronomy and Astrophysics, Tata Institute of
  Fundamental Research, Colaba, Mumbai-400005, India
    \and
  Astrophysikalisches Institut Potsdam, An der Sternwarte 16, D-14482 Potsdam, Germany
  \and
   University of California San Diego, Center for Astronomy and Space
Sciences, 9500 Gilman Drive \#0425, La Jolla, CA, 92093-0424, USA
  \and
Mullard Space Science Laboratory, University College London, Holmbury St. Mary, RH5 6NT Dorking, UK
 \and
  Universit\'e de Toulouse;
   UPS-OMP; IRAP, Toulouse, France
        \and 
         CNRS; IRAP; 9 avenue du Colonel Roche, BP 44346,
		 F-31028 Toulouse Cedex 4, France}
   \date{Received xx.xx.2011; accepted xx.xx.2012}

  \abstract
   {}
   {Reliable timing calibration is essential for the accurate comparison of \emph{XMM-Newton} light curves with those from other observatories, to ultimately use them to derive precise physical quantities. The \emph{XMM-Newton} timing calibration is based on pulsar analysis. However, as pulsars show both timing noise and glitches, it is essential to monitor these calibration sources regularly. To this end, the \emph{XMM-Newton} observatory performs observations twice a year of the Crab pulsar to monitor the absolute timing accuracy of the EPIC-pn camera in the fast Timing and Burst modes. We present the results of this monitoring campaign, comparing \emph{XMM-Newton} data from the Crab pulsar (PSR B0531+21) with radio measurements. In addition, we use five pulsars (PSR J0537-69, PSR B0540-69, PSR B0833-45, PSR B1509-58 and PSR B1055-52) with periods ranging from 16 ms to 197 ms to verify the relative timing accuracy.}
{We analysed 38 \emph{XMM-Newton} observations (0.2-12.0 keV) of the Crab taken over the first ten years of the mission and 13 observations from the five complementary pulsars. All the data were processed with the \emph{SAS}, the \emph{XMM-Newton Scientific Analysis Software}, version 9.0. Epoch folding techniques coupled with $\chi^{2}$ tests were used to derive relative timing accuracies. The absolute timing accuracy was determined using the Crab data and comparing the time shift between the main X-ray and radio peaks in the phase folded light curves.}
   {The relative timing accuracy of \emph{XMM-Newton} is found to be better than $10^{-8}$. The strongest X-ray pulse peak precedes the corresponding radio peak by 306$\pm$9 $\mu$s, which is in agreement with other high energy observatories such as \emph{Chandra}, \emph{INTEGRAL} and \emph{RXTE}. The derived absolute timing accuracy from our analysis is $\pm$48 $\mu$s.}
   {}

\color{black}
\authorrunning {A. Martin-Carrillo et al.}
\titlerunning  {The rel. and abs. timing of \emph{XMM-Newton} derived from the Crab and other pulsars. } 
\maketitle
%

\section{Introduction}
A reliable timing calibration is essential for all timing data analyses 
and the physics derived from those.
Irregularities in the spacecraft time correlation, the on-board instrument
oscillators or data handling unit and the ground processing and data analysis 
software can lead to errors in relative and absolute information pertaining to the timing 
behaviour of astrophysical objects. 
The timing of the \emph{XMM-Newton} observatory is evaluated using 
\emph{XMM-Newton's} EPIC-pn camera that has been extensively ground calibrated
with respect to relative timing, but due to a limited calibration time budget, the end-to-end system for absolute timing was never checked on the ground.
The relative timing for fast sources like
the Crab was expected to have an accuracy of $\Delta\,P/P\,\lesssim\,10^{-8}$ before launch. For the absolute timing
a requirement of $\Delta\,T\,\lesssim\,1 \,ms$ was given.

\emph{XMM-Newton} \citep{jansen01} was launched in December 1999 with an Ariane 5 rocket from French Guayana. It operates six
instruments in parallel on its 48\,hour highly elliptical orbit:
three Wolter type~1 telescopes, with 58 nested mirror shells each,
focus X-ray photons onto the three X-ray instruments of the EPIC (European Photon Imaging Camera)
\citep{strueder01,turner01} and the two Reflecting Grating Spectrometers
\citep[RGS;][]{herder01}.  In addition, a 30\,cm Ritchey Chr\'etien
optical telescope, the Optical Monitor, is used for optical observations
 \citep[OM;][]{mason01}.  EPIC consists of three
cameras: the two EPIC-MOS cameras use Metal-Oxide Semiconductor CCDs
as X-ray detectors, while the EPIC-pn camera is equipped with a
pn-CCD. All three have been especially developed for \emph{XMM-Newton}
\citep{pf99,me99,turner01}.

In this paper we determine the relative timing accuracy of \emph{XMM-Newton's} EPIC-pn camera using all 
available observations of the Crab pulsar in combination with other isolated pulsars in order to extend our 
analysis to a broader variety of sources. Preliminary results on the relative timing accuracy of \emph{XMM-Newton} 
using the Crab pulsar and the other pulsars can be found in \citet{caballero}.

In this work we use only the Crab pulsar X-ray observations to determine the  absolute timing accuracy.  However, as this is done in reference to radio timing, it is limited to the accuracy of the radio ephemerides.

We also see the paper as a summary of "how to perform" relative and absolute timing analysis with 
\emph{XMM-Newton} and what timing accuracy the user can expect for different targets.

This paper is organised as follows. In Sect.~\ref{sec:targets} we give a description of the targets used for the timing evaluation, followed by some technical comments on 
our data analysis in Sect.~\ref{sec:ana}. The relative and absolute timing results are presented in 
Sect.~\ref{sec:relt} and Sect.~\ref{sec:abst}. A short description of the 
\emph{XMM-Newton's} EPIC-pn camera is given in Appendix~\ref{sec:xmm}.

\section{Observations}
\label{sec:targets}
All pulsars used in our analysis are isolated. We concentrated primarily on the Crab pulsar (PSR B0531+21) as radio ephemerides 
are provided monthly by the Jodrell Bank Observatory\footnote{\texttt{http://www.jb.man.ac.uk} \citep{lyne}}.
The other pulsars have been chosen to include a range of periods and pulse profiles, with which to check the relative timing. Some of these pulsar observations were reported by 
\citet{becker} as a summary of first results from \textit{XMM-Newton}.

Tables~\ref{tab:periods} and  \ref{tab:periodso} summarise the data used and the results obtained from all the Crab observations studied and all the other pulsars respectively. 
Column~1 gives the observation ID (OBSID) used for identifying \emph{XMM-Newton} observations, followed 
by the satellite revolution ("Rev.") in which the observation was done, the data mode, and the filter used.  Column~5 indicates whether the observation is affected by telemetry gaps (due to a full science buffer), and column~6 
gives information on time jumps during the observation (see the footnote of the table for explanation). 
Column~7 lists the start times ("Epoch") of the observations in MJD, followed by the exposure ("Obs. Time") in ks.
Columns~9 and 10 list the pulse periods of the Crab pulsar in the radio at the time of the \emph{XMM-Newton} 
observations (interpolated using the information provided by the Jodrell Bank Observatory) and the measured
X-ray period, respectively. Red. $\chi^{2}$ (column~11) gives the reduced $\chi^{2}$ values found at the maximum
of the respective $\chi^{2}$ distribution of the period search (the number of degrees of freedom, \textit{dof}, was always 100 for the Crab pulsar), and
"FWHM" is the full width at half maximum of the $\chi^{2}$ distribution.
$\Delta$P/P is the relative difference
between the radio and the X-ray period (Eq.~\ref{eq:rel} in Sect.~\ref{subsec:rel}). The "Phase Shift" (last column) shows 
the measured time shift of the main peak in the pulse profile between the X-ray and radio profiles, 
as explained in Sect.~\ref{subsec:abs}. All uncertainties given are at the 1 $\sigma$ (68\%) level.
The ephemerides of all the targets used in the analysis are shown in Table~\ref{tab:radioeph}.
Fig.~\ref{alllight} shows the pulse profiles for all the pulsars analysed in this paper.

\subsection{The main XMM-Newton timing monitoring source: PSR B0531+21 (The Crab pulsar)}
Since the discovery of the Crab Pulsar \citep{Staelin Reifenstein}, the Crab has been one of the best studied 
objects in the sky and it remains one of the brightest X-ray sources regularly observed.  As a standard candle for instrument calibration, 
the 33\,ms Crab pulsar has been repeatedly studied (monitored) by many astronomy missions in almost 
every energy band of the electromagnetic spectrum. However, recent analysis presented by \citet{wilson-hodge}
showed that the flux of the Crab is not constant on long timescales at high energies. These flux variations seem to be related to the 
nebula and correspond to a flux drop of $\sim$7 \% (70 mCrab) over two years (2008-2010). 
This might affect the status of the Crab as a standard candle in the future.

In the X-ray regime its pulse profile exhibits a double peaked structure with a phase separation of 0.4 between 
the first (main) and the second peak. X-ray emission at all phases, including the pulse minimum, was discovered by 
\citet{tennant01} using the \textit{Chandra} observatory. Measurements of X-ray to radio delays between the arrival 
times of the main pulse in each energy range of the Crab pulsar have been reported using all high-energy 
instruments aboard \textit{INTEGRAL} \citep{kuiper 2003} and \textit{RXTE} \citep{rots 2004}.  The time 
delays were determined to be 280$\pm$40 $\mu$s and 344$\pm$40 $\mu$s respectively.

The Crab pulsar has been observed bianually to monitor the timing capabilities of \emph{XMM-Newton}. Over the 
years an observation strategy has been established that makes very efficient use of the limited calibration time budget.
\emph{XMM-Newton} generally observed the Crab pulsar three times per orbit for 5 ks: at the beginning, in the middle, 
and the end of that orbit. These campaigns were carried out in spring and autumn when
\emph{XMM-Newton} has a different location in its orbit with respect to the Sun-Earth system. This guarantees
the monitoring of the dependency of the timing with respect to \emph{XMM-Newton's} orbital position. Eventual irregularities in relative timing with respect to the orbital position could then be identified. A total of 38 observations with exposure times between 2\,ks and 40\,ks have been analysed in this paper. See Table~\ref{tab:periods} for details of these observations.

\subsection{Other useful pulsars for relative timing analysis}
\subsubsection{PSR J0537-69}
PSR J0537-69 is a young pulsar, about 5000 years in age, located in the Large Magellanic Cloud. It is embedded in the supernova 
remnant N157B and is considered to be the oldest known Crab-like pulsar. It is a very fast-spinning pulsar with a period of 
16\,ms, discovered by \citet{marshall98} using \textit{RXTE}. No significant radio signal above a 5$\sigma$ threshold has been detected from the pulsar 
 \citep{crawford}. In the X-ray energy range, \textit{RXTE} has monitored PSR J0537-69 for seven years \citep{marshall 2004,middleditch}, providing a complete study of the behaviour of the pulsar. \citet{middleditch} reported 23 sudden increases in frequency, called \textit{glitches} and present in most of young pulsars. 
Due to this highly irregular activity (a glitch every $\sim$4 months) a contemporaneous ephemeris is important. Its pulse profile in the X-ray regime is characterized by a single narrow peak. 

See Table~\ref{tab:periodso} for details of the observations. 
Our 36\,ks observation coincides with the \textit{RXTE} monitoring campaign presented by \citet{middleditch} and a good 
ephemeris was therefore guaranteed.

\subsubsection{PSR~B0540-69}
This young pulsar ($\sim$1500 years) was discovered in soft X-rays by 
\citet{seward} with a period of 51\,ms, in the field of the Large Magellanic Cloud and it is considered to be a Crab-like pulsar. Its pulse 
shape does not appear to change significantly from optical to hard X-rays \citep{deplaa}.
The pulsed radio emission was discovered in late 1989 appearing as a faint source \citep{manchester} and presenting 
a complex profile, very different from the simple sinusoidal one seen in X-rays (Fig.\ref{alllight}).
A glitch was reported by \citet{zhang} before the \emph{XMM-Newton} observations
and confirmed by \citet{livingstone 2005b} using a 7.6\,year \textit{RXTE} campaign. The glitch 
activity of PSR~B0540-69 is known to be less than that of the Crab pulsar \citep{livingstone 2005b} but the 
presence of considerable timing noise was reported by \citet{cusumano} using \textit{ASCA}, \textit{BeppoSAX} 
and \textit{RXTE} observations made over a time interval of about 8 years. Therefore, despite the low glitch 
activity, long extrapolations of the ephemeris might not be reliable.
See Table~\ref{tab:periodso} for details of the observations.

\subsubsection{PSR~B0833-45 (Vela pulsar)}
The Vela pulsar with a period of 89\,ms was discovered by \citet{large} and it is associated with a supernova remnant. It is, with the Crab pulsar, one of the most active young/middle-age pulsars known, showing
regular glitches. These glitches have been intensively studied 
for the Vela pulsar, where a dozen events in different 
energy ranges have been recorded and analysed over the past three decades \citep{helfand,dodson}. Due to these important irregularities close radio ephemerides are needed.

No Vela timing mode observations have been performed with \emph{XMM-Newton}, but since the period is 
89\,ms the data in Small Window mode (time resolution of 5.7 ms) can be used for our purposes. Thus we have analysed 
the four observations listed in Table~\ref{tab:periodso}.

\subsubsection{PSR B1509-58}
This young pulsar ($\sim$1700 years) is one of the most energetic pulsars known and has a pulse period of 
$\sim$151 ms. It is associated with the supernova remnant G320.4-1.2 and it has been well studied in all wavelengths 
since it was discovered in the soft X-ray band using \emph{Einstein} \citep{seward82}. The pulse profile in X-rays of appears to be much broader than in radio, changing from a narrow peak shape into a more sinusoidal 
shape at high energies. Monitored by \textit{RXTE} since its launch, and covering a 21 year time interval and in conjunction with radio data 
from the \textit{MOST} and \textit{Parkes} observatories, a detailed timing study has been carried out \citep{livingstone 2005a}, but no glitch was found in the entire data sample. This result makes PSR~B1509-58 probably the only known young pulsar that does not present any glitches over long periods of time. This property means that it is well adapted to extrapolation over long time intervals and useful for absolute timing analyses \citep{rots 1998}. See Table~\ref{tab:periodso} for details of the observations.

\subsubsection{PSR B1055-52}
PSR B1055-52, one of the Three Musketeers together with PSR B0656+14 and 
Geminga, is a middle-aged pulsar with a period of 197 ms. It was discovered by \citet{vaughan} but it was only in
1983 that X-ray emission was first detected by \citet{cheng} using the {\em Einstein} Observatory. \citet{oegelman} detected sinusoidal pulsations in X-rays up to 2.4 keV. More recently, \citet{deluca} showed  using \emph{XMM-Newton} data that the pulsed emission is detectable up to 6 keV. Most middle-aged pulsars like PSR B1055-52 show reduced timing noise and fewer glitches compared to younger ones. See Table~\ref{tab:periodso} for details of the observations. Results concerning these data have been originally published by \citet{deluca}.

\begin{figure} [!ht]
	
\centering

\resizebox{\hsize}{!}{\includegraphics[angle=0, bb=47 150 550 780, clip=,]{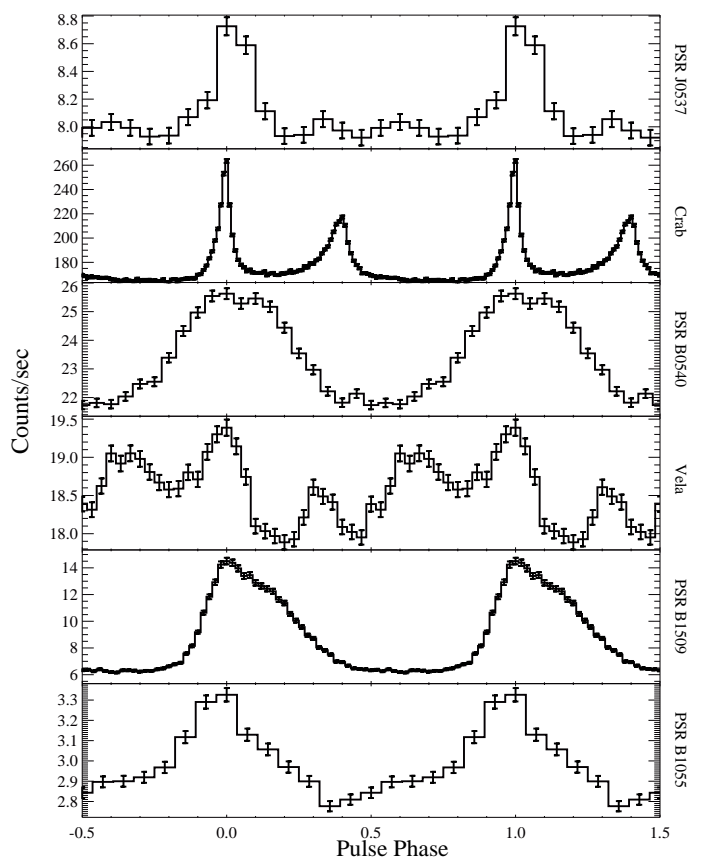}}
\caption{XMM-Newton pulse profiles of the different pulsars analysed. From top to bottom: PSR~J0537-69 (obs. ID: 0113020201), 
the Crab pulsar (obs. ID: 0122330801), PSR~B0540-69 (obs. ID: 0125120201), the Vela pulsar (obs. ID: 0111080201), PSR~B1509-58 (obs. ID: 0312590101) and 
PSR~B1055-52 (obs. ID: 0113050201) with periods of 16 ms, 33 ms, 51 ms, 89 ms, 151 ms and 197 ms, respectively. 
The energy band  in all cases is 0.2-12 keV.}
\color{black}
\label{alllight}
\end{figure}
\begin{landscape}
\begin{table} [h]
\caption{Individual observations from the Crab monitoring. (A description of the columns is given in the text.)}
\label{tab:periods}
\begin{tabular}{lrrccccccccccccccccc}
\hline
\hline
       \noalign{\smallskip}
   OBSID  & Rev. & Mode & Filter & Counting mode & Time Jump & Epoch  & Obs. Time  & Radio period & X-ray period & Red. $\chi^{2}$ & FWHM    &   $\Delta$P/P  & Phase Shift\\
          &      &      &        &               &       & [MJD]  & [ks]       &    [ms]      &   [ms]       &                & $10^{-8}s$ &  $10^{-8}$    & [$\mu$s]  \\
   
            \noalign{\smallskip}
            \hline
            \noalign{\smallskip}

0122330801 & 56   & TI & T & NO  & S & 51632.83 & 22.58 & 33.5083819072 & 33.5083817(1) &  433 & 1.37  & -0.6$\pm$0.4  & -148$\pm$2 \\
0135730701 & 234  & B  & T & NO  & N & 51988.64 & 10.00 & 33.5213091402 & 33.5213091(5) &  197 & 3.24  & 0.02$\pm$0.9  & -337$\pm$4 \\
0153750201 & 411  & B  & M & YES & N & 52340.68 & 4.63  & 33.5341004722 & 33.5341001(7) &   79 & 6.78  & -0.9$\pm$2    & -287$\pm$4 \\
0153750301 & 411  & B  & M & YES & N & 52341.29 & 9.00  & 33.5341224876 & 33.5341223(9) &  143 & 3.52  & -0.3$\pm$1    & -296$\pm$4 \\
0153750401 & 411  & TI & T & YES & N & 52341.42 & 9.00  & 33.5341272691 & 33.5341278(9) &  269 & 3.57  &  2$\pm$1	   & -217$\pm$2 \\
0153750501 & 411  & B  & M & YES & N & 52341.84 & 9.00  & 33.5341426021 & 33.5341426(5) &  136 & 3.55  & 0.2$\pm$1	   & -257$\pm$4 \\
0160960201 & 698  & TI & T & YES & N & 52913.51 & 28.26 & 33.5549129334 & 33.5549128(4) &  912 & 1.58  & -0.3$\pm$0.5  & -174$\pm$2 \\
0160960301 & 700  & TI & T & YES & N & 52918.30 & 10.17 & 33.5550871157 & 33.5550881(2) &  313 & 4.58  & 3$\pm$1	   & -292$\pm$3 \\
0160960401 & 874  & B  & T & NO  & N & 53264.10 & 14.72 & 33.5676516600 & 33.5676516(5) &  250 & 2.20  & -0.04$\pm$0.7  &  603$\pm$3 \\
0160960601 & 874  & B  & T & NO  & N & 53265.57 & 3.98  & 33.5677052394 & 33.5677050(1) &   68 & 8.01  & -0.7$\pm$2    & 8839$\pm$5 \\
0160960701 & 955  & B  & T & YES & N & 53425.72 & 8.20  & 33.5735256348 & 33.5735259(2) &  133 & 3.86  & 0.9$\pm$1	   & -334$\pm$3 \\
0160960801 & 955  & B  & T & YES & N & 53426.27 & 5.00  & 33.5735454128 & 33.5735452(4) &   80 & 6.52  & -0.5$\pm$2    & -336$\pm$5 \\
0160960901 & 955  & B  & T & YES & N & 53426.77 & 8.20  & 33.5735638700 & 33.5735642(0) &  127 & 3.90  & 1$\pm$1	   & -357$\pm$4 \\
0160961001 & 1048 & TI & T & YES & N & 53612.63 & 4.99  & 33.5803179048 & 33.580317(7)  &  178 & 11.18 & -0.6$\pm$3    & -257$\pm$3 \\
0160961101 & 1049 & B  & T & YES & N & 53613.14 & 5.00  & 33.5803366674 & 33.5803368(8) &   59 & 6.60  & 0.6$\pm$2	   & -335$\pm$4 \\
0160961201 & 1049 & TI & T & YES & N & 53613.92 & 4.99  & 33.5803650194 & 33.580364(2)  &  164 & 10.25 & -3$\pm$3    & -286$\pm$2 \\
0160961301 & 1140 & B  & T & NO  & N & 53794.72 & 5.00  & 33.5869348775 & 33.5869347(5) &   90 & 6.77  & -0.4$\pm$2    & -284$\pm$4 \\
0160961401 & 1140 & TI & T & YES & N & 53795.44 & 4.99  & 33.5869607624 & 33.5869607(3) &  262 & 6.18  & -0.1$\pm$2    & -209$\pm$2 \\
0160961501 & 1140 & B  & T & NO  & N & 53796.09 & 6.70  & 33.5869844395 & 33.5869844(8) &  129 & 4.47  & 0.1$\pm$1	   & -305$\pm$4 \\
0312790101 & 1138 & B  & T & YES & E & 53791.22 & 42.50 & 33.5868077617 & 33.5868078(5) &  422 & 1.25  &  0.3$\pm$0.4  & -268$\pm$2 \\
0312790201 & 1138 & B  & T & NO  & N & 53791.75 & 5.03  & 33.5868269683 & 33.5868267(9) &   72 & 6.46  & -0.5$\pm$2    & -349$\pm$5 \\
0312790401 & 1138 & B  & T & NO  & N & 53791.95 & 7.80  & 33.5868339621 & 33.5868344(5) &   52 & 4.19  & 1$\pm$1	   & -305$\pm$6 \\
0412590101 & 1249 & B  & T & NO  & N & 54012.06 & 6.40  & 33.5948315461 & 33.5948316(2) &  111 & 4.82  &  0.2$\pm$1	   & -369$\pm$4 \\
0412590201 & 1249 & TI & T & YES & N & 54012.72 & 5.04  & 33.5948556935 & 33.5948540(4) &  282 & 9.77  & -5$\pm$3    & -284$\pm$2 \\
0412590301 & 1249 & B  & T & NO  & N & 54013.38 & 8.85  & 33.5948794078 & 33.5948792(9) &  148 & 3.39  & -0.4$\pm$1    & -389$\pm$3 \\
0412590601 & 1325 & TI & T & YES & N & 54164.32 & 5.04  & 33.6003636968 & 33.6003640(7) &  241 & 6.14  &  1$\pm$2   & -524$\pm$2 \\
0412590701 & 1325 & B  & T & YES & N & 54164.98 & 8.85  & 33.6003876860 & 33.6003877(0) &  156 & 3.50  &  0.03$\pm$1	   & -601$\pm$3 \\
0412591001 & 1414 & B  & T & NO  & N & 54341.11 & 13.20 & 33.6067862758 & 33.6067861(8) &  220 & 2.33  & -0.3$\pm$0.7  & -336$\pm$3 \\
0412591101 & 1414 & TI & T & YES & N & 54341.96 & 5.00  & 33.6068170822 & 33.6068166(3) &  308 & 8.62  & -1$\pm$3    & -303$\pm$3 \\
0412591201 & 1414 & B  & T & NO  & S & 54342.40 & 13.41 & 33.6068330929 & 33.6068330(5) &  228 & 2.40  & -0.1$\pm$0.7  & -360$\pm$3 \\
0412591401 & 1504 & B  & T & YES & N & 54520.72 & 21.70 & 33.6133108411 & 33.6133108(9) & 394	  & 1.45  &  0.1$\pm$4    & -357$\pm$2 \\
0412591501 & 1504 & TI & T & YES & N & 54521.46 & 7.64  & 33.6133378231 & 33.6133377(7) &	 521 & 4.53  & -0.1$\pm$1   & -282$\pm$2 \\
0412591601 & 1504 & B  & T & YES & N & 54521.85 & 20.20 & 33.6133519405 & 33.6133519(9) & 382	  & 1.51  &  0.1$\pm$0.4  & -363$\pm$2 \\
0412591901 & 1600 & B  & T & YES & N & 54712.22 & 5.80  & 33.6202664839 & 33.6202667(5) &	 133 & 5.50  & 0.8$\pm$2   & -282$\pm$3 \\
0412592001 & 1600 & TI & T & YES & N & 54712.78 & 6.04  & 33.6202865236 & 33.6202880(2) &	 544 & 7.19  & 4$\pm$2    & -207$\pm$2 \\
0412592101 & 1600 & B  & T & YES & N & 54713.35 & 15.39 & 33.6203076395 & 33.6203076(3) & 365  & 1.98  & -0.02$\pm$0.6  & -263$\pm$2 \\
0412592401 & 1687 & B  & T & YES & N & 54885.70 & 18.90 & 33.6265670615 & 33.6265671(0) & 355  & 1.67  & 0.1$\pm$0.5  & -379$\pm$2 \\
0412592501 & 1687 & TI & T & YES & N & 54886.52 & 4.54  & 33.6265967322 & 33.6265966(8) &	 341 & 6.78  & -0.1$\pm$2   & -305$\pm$2 \\

            \noalign{\smallskip}
            \hline

\end{tabular}
\tablefoot{
N: no time jump,
S: Time jump corrected by the SAS,
E: Time jump not corrected by the SAS, data had to be partly excluded from the analysis. TI: Timing Mode,
B: Burst Mode,
M: Medium Filter,
T: Thick Filter. Epoch: the epoch of the 1st  \emph{XMM-Newton} event.
Radio period: the radio period extrapolated to the \emph{XMM-Newton} epoch. The accuracy of the radio periods can be considered to be good to $10^{-13}$, which is almost negligible compared to the X-ray errors of the period.
}

\end{table}

\end{landscape}


\begin{landscape}
\begin{table} [h]
\caption{Individual observations of the other objects.}
\label{tab:periodso}
\begin{tabular}{lrrccccccccccccccccccc}
\hline
\hline
       \noalign{\smallskip}
  
  Object & OBSID  & Rev. & Mode & Filter & Epoch  & Obs. Time  & Reference period &  Reference Epoch & X-ray period & Red. $\chi^{2}$ & FWHM       & $\Delta$P/P  \\
         &        &      &      &        & [MJD]  & [ks]       &     [ms]         &   [MJD]          &      [ms]    &               & $10^{-8}s$ &  $10^{-8}$   \\
             \noalign{\smallskip}
            \hline
            \noalign{\smallskip}

PSR J0537$^{X}$ & 0113020201 & 357  & TI & M  & 52232.95  & 35.93  &  16.1229802136  & 52260.40296  & 16.1229802(3)   & 24   &   0.38  &$-0.03\pm2$  \\
PSR B0540$^{X}$ & 0125120201 & 85   & TI & M  & 51691.54  & 17.05  &  50.5192855014  & 51686.20757   & 50.519284(6)  & 99   &  12.60  &$ -2.40\pm12$ \\
PSR B0540$^{R}$ & 0125120201 & 85   & TI & M  & 51691.54  & 17.05  &  50.5194431419  & 52857.86600   & 50.519284(6)  & 99   & 12.60  &$ -31.44\pm12$\\
PSR B0540$^{X}$ & 0413180201 & 1248 & TI & M  & 54010.06  & 12.64  &  50.6160286612  & 52857.86600  & 50.615225(9)  & 112  &  17.34  &$ -1580\pm17$ \\
PSR B0540$^{X}$ & 0413180301 & 1248 & TI & M  & 54010.79  & 15.24  &  50.6160594083  & 52857.86600  & 50.615252(7)  & 136  &  14.75  &$ -1594\pm15$ \\
Vela$^{R}$     & 0111080101 & 180  & SW & M  & 51880.00  & 37.90  & 89.3318630836   & 52408.00000      & 89.331848(3)   &  18  &   7.90  &$ -17.35\pm3$\\
Vela$^{R}$     & 0111080201 & 180  & SW & M  & 51880.51  & 58.60  & 89.3318685573   & 52408.00000     & 89.331853(2)   &  25  &   6.01  &$ -17.84\pm2$\\
Vela$^{R}$     & 0111080301 & 180  & SW & M  & 51881.23  & 1.94   & 89.3318762404   & 52408.00000       & 89.33181(3)   &   2  & 81.00  &$ -79.22\pm30$\\
Vela$^{R}$     & 0153951401 & 1169 & SW & M  & 53852.54  & 120.87 & 89.3531575271   & 53193.00000    & 89.352786(3)   &  17  &  8.47  &$-415.46\pm3$\\
PSR B1509$^{R}$ & 0128120401 & 137  & TI & M  & 51794.26  & 9.60   & 151.1145869352  & 50352.00000   & 151.11357(1)  & 198  & 190.19 &$-673.21\pm25$\\
PSR B1509$^{R}$ & 0312590101 & 1136 & TI & M  & 53786.98  & 31.95  & 151.3773132754  & 53385.00000  & 151.37723(1)  & 694  &  58.35  &$ -52.25\pm8$\\
PSR B1055$^{R}$ & 0113050101 & 186  & TI & M  & 51892.93  & 19.90  & 197.1118094698  & 50256.00000    & 197.1118(1)  &  30  & 151.09  &$  12.83\pm55$\\
PSR B1055$^{R}$ & 0113050201 & 187  & TI & M  & 51893.73  & 52.70  & 197.1118098756  & 50256.00000    & 197.11183(5)  &  34  &  65.81  &$  11.09\pm24$\\

            \noalign{\smallskip}
            \hline
	    \hline

\end{tabular}
\tablefoot{
M: Medium Filter; TI: Timing Mode; SW: Small Window Mode. $^{(R)}$: reference ephemeris from radio data; $^{(X)}$: reference ephemeris from RXTE data.
}
\end{table}
\end{landscape}

\section{Data analysis}
\label{sec:ana}
The data sets were processed using the \emph{XMM-Newton} Scientific Analysis Software, \textit{SAS 9.0} \citep{gabriel}.  
Event times were corrected to the solar system barycentre using the SAS tool \texttt{barycen}.

\subsection{Relative timing data analysis}
\label{subsec:rel}
We define the relative timing accuracy as the difference between the period measured with \emph{XMM-Newton} and the period measured at radio wavelengths evaluated at the epoch of the X-ray observations. This difference is normalised to the pulse period measured in radio.

\begin{equation}
\label{eq:rel}
\centering \texttt{Rel. timing}:=\dfrac{P_{\texttt{X-ray}}(T_{\texttt{X-ray}})-P_{\texttt{radio}}(T_{\texttt{X-ray}})}{P_{\texttt{radio}}(T_{\texttt{X-ray}})}=\dfrac{\Delta P}{P}
\end{equation}
where 
\begin{itemize}

\item[ ] $P_{\texttt{X-ray}}:$  period derived from \emph{XMM-Newton} 
\item[ ] $P_{\texttt{radio}}:$  period extrapolated from radio ephemeris  
\item[ ] $T_{\texttt{X-ray}}:$  time of the first X-ray event of the \emph{XMM-Newton} observation [MJD]

\end{itemize}

We determined the period of the Crab pulsar in X-rays using the epoch folding software
 \texttt{XRONOS}\footnote{\texttt{XRONOS} is part of the HEARSAC software (\texttt{http://heasarc.gsfc.nasa.gov}).}. 
The closest available radio ephemeris (supplied by the Jodrell Bank Crab Pulsar Monthly Ephemeris) before 
and after the X-ray observation were used to interpolate the radio period \textit{P} for the time of the first 
X-ray event of the \emph{XMM-Newton} observation in MJD. The interpolated radio periods 
are then used as an initial trial value for the epoch folding. The period derivative $\dot{P}$ 
provided by Jodrell Bank is taken into account when doing the folding of the X-ray data. All relevant initial and final values are listed in Tables~\ref{tab:periods}, \ref{tab:periodso}, \ref{tab:epfolding} and \ref{tab:radioeph}.
All X-ray pulse profiles shown in Fig.~\ref{alllight} have been produced using the best fit X-ray period.

The detailed steps of our data reduction are presented below in order to provide 
an example for proper \emph{XMM-Newton} relative timing data analysis.

\begin{enumerate}

\item calibrate the \emph{XMM-Newton} event list using the \textit{SAS} routine 
\texttt{epproc}\footnote{command line set up of 
\texttt{epproc}: timing=YES burst=YES srcra=83.633216667 srcdec=22.014463889 withsrccoords=yes}

\item perform barycentre correction using precise coordinates with the \textit{SAS} routine 
\texttt{barycen}\footnote{command line set up of \texttt{barycen}: 
withtable=yes table='bary.ds:EVENTS' timecolumn='TIME' withsrccoordinates=yes srcra='83.633216667' srcdec='22.014463889' processgtis=yes time=0}

\item extract source\footnote{Detector coordinates used in the extraction process a) timing mode: 
(RAWX,RAWY) IN box(35,101,12,100,0), b) burst mode: (RAWX,RAWY) IN box(35.,71.5,20,70,0)} 

\item extrapolate the radio ephemeris

\item period search using \texttt{efsearch} from \texttt{XRONOS} (see Table~\ref{tab:epfolding}) which gives the $\chi^{2}$ against the period

\item period determined through a weighted mean of all values within 65$\%$ of the \texttt{efsearch} $\chi^{2}$  maximum 

\end{enumerate}

\begin{table} [ht!]
\caption{Settings for the epoch folding using \texttt{efsearch}.}
\label{tab:epfolding}
\begin{tabular}{lrrcccccc}
\hline
\hline
       \noalign{\smallskip}
   Object  & nper  & nphase & dres & \.P  \\
  
            \noalign{\smallskip}
            \hline
            \noalign{\smallskip} 
           PSR J0537  &   5000   & 15  &  $10^{-10}$  &  5.1815 $10^{-14}$    \\
	   Crab      &   5000   & 100  &  $10^{-10}$  &  0.4205 $10^{-12}$    \\
	   PSR B0540  &  10000   &  20  &  $10^{-10}$  &  4.78907 $10^{-13}$   \\
	   Vela      &   2000   &  30  &  $10^{-9} $  &  1.25008  $10^{-13}$  \\
	   PSR B1509  &  50000   &  50  &  $10^{-10}$  &  1.53085 $10^{-12}$   \\
	   PSR B1055  &  40000   &  14  &  $10^{-10}$  &  5.8354873 $10^{-15}$ \\
	    \noalign{\smallskip}
            \hline

\end{tabular}
\newline 
\tablefoot{\emph{nper}: number of periods over which the search is carried out; \emph{nphase}: resolution of the trial folded light curves (number of bins in each period); \emph{dres}: resolution for the period search in seconds; \emph{\.P}: period derivative.
}
\end{table}

\begin{table*}[!ht]
\caption{Radio and RXTE ephemerides used in the analysis.}
\label{tab:radioeph}
\begin{tabular}{lrrccccccc}
\hline
\hline
       \noalign{\smallskip}
   Object&  RA   & Dec   & epoch $E_{0}$  & ${\nu(E_{0})}$ & $\dot{\nu}(E_{0})$ & $\ddot{\nu}(E_{0})$ \\
         &(J2000)&(J2000)&    [MJD]       &  [mHz]         & [x10$^{-10}$ Hz$s^{-1}$]          &  [x10$^{-21}$ Hz$s^{-2}$]       \\
            \noalign{\smallskip}
            \hline
            \noalign{\smallskip} 
       PSR J0537$^{1}$  &05:37:47.20 &-69:10:23.00 & 52260.40296   & 62.02279999(0)  &   -1.993398(1) & 6.80(0)  \\
	   Crab$^{2}$          &05:34:31.97 &22:00:52.07   & --   & -- & --   & --   \\
	   PSR B0540$^{3}$ &05:40:11.22 &-69:19:54.98 &51686.20757        & 19.794507(5)                   &  -1.881021(0)  & 3.79(9)  \\
       PSR B0540$^{4}$ &05:40:11.22 &-69:19:54.98 &52857.86600         &  19.77552961(8)             &  -1.87288(3)  & 4.30(2)  \\
       PSR B0540$^{4}$ &05:40:11.22 &-69:19:54.98 &52857.86600         &  19.775529611(3)            &  -1.872853(1)  & 4.18(1)  \\
	   Vela $^{5}$         &08:35:20.61 &-45:10:34.87  &  52408.00000       & 11.193503640388(1)       &  -0.156027(3)  &  0.64(1) \\
	   Vela$^{6}$          &08:35:20.61 &-45:10:34.87  &53193.00000        & 11.192447207118304(3) & -0.155502(8) & 0.52(7) \\
	   PSR B1509$^{7}$ &15:13:55.62 &-59:08:09.00  & 50352.00000       &  6.625918674074(4)         & -0.673579(0)   & 1.95(0) \\
	   PSR B1509$^{8}$ &15:13:55.62 &-59:08:09.00 & 53385.00000        & 6.608333867737907(3)    &   -0.66852(2) & 1.91(4) \\
	   PSR B1055$^{9}$ &10:57:58.84 &-52:26:56.30 & 50256.00000        & 5.073283989285(9)          &   -0.015019(5) & 3.24(0) \\
	    \noalign{\smallskip}
            \hline
\end{tabular}
\tablebib{(1)~\citep{middleditch}; (2)~Jodrell Bank Crab Pulsar Monthly Ephemeris; (3)~\citep{cusumano}; (4)~\citep{johnston}; (5)~\citep{romani}; (6)~\citep{dodson}; (7)~\citep{livingstone 2005a}; (8) Parkes Observatory (private communication); (9) Princeton Database.  
}
\end{table*}

The number of phase bins per period (nphase) in each pulse profile was chosen such that the count rate  
uncertainties in each bin (determined using  the Poisson error on the count rate per
bin normalised by the bin size) are, on average, not bigger than approximately 10\% of the total count rate variation in the 
pulse profile of the shortest observation for each pulsar. This value, determined for each pulsar, is used for all the 
observations of that object. In this way the signal to noise in each bin is sufficient to reliably determine any 'smearing out' of the pulse profile due to the use of an inaccurate period/ephemeris, essential for determining the relative timing precision, as described in Sect. \ref{sec:relt}. 

\subsection{Absolute timing data analysis}
\label{subsec:abs}

The  \emph{XMM-Newton} EPIC-pn absolute timing accuracy was determined using 
only observations of the Crab. The ephemeris  (epoch, \textit{P}, $\dot{P}$, $\ddot{P}$) 
of the nearest radio observation from the Jodrell Bank Observatory was used as a reference to obtain the phase shift between the time of arrival of the main peak in the X-ray profile and the time of arrival of the main peak in the radio profile, as described in Eq.~\ref{eq:abs}. The phase shift was then multiplied by the corresponding X-ray period found during the relative timing analysis, as shown in Table~\ref{tab:periods}.

\begin{equation}
\label{eq:abs}
\centering \texttt{Phase Shift [$\mu$\,s]} := {T_{0}}_{\texttt{X-ray}}-{T_{0}}_{\texttt{radio}}
\end{equation}
where 
\begin{itemize}

\item[ ] ${T_{0}}_\texttt{{X-ray}}:$  Time of arrival of the main peak of the X-ray profile
\item[ ] ${T_{0}}_{\texttt{radio}}:$  Time of arrival of the main peak of the radio profile

\end{itemize}

\begin{figure}[ht]
\begin{center}
\resizebox{\hsize}{!}{\includegraphics[angle=90, bb=40 40 530 740, clip=,]{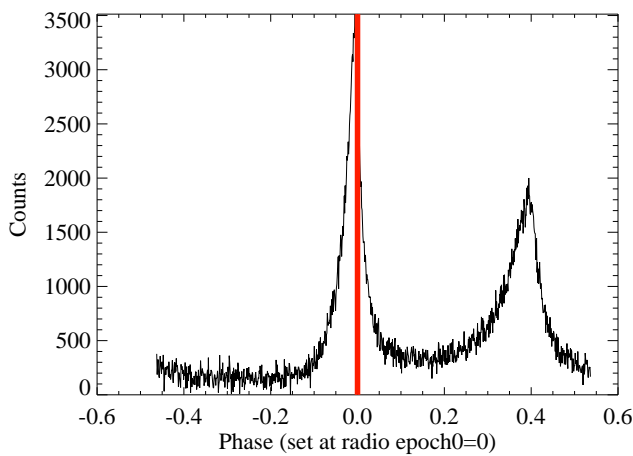}}
\caption{Crab pulse profile (obs. ID: 0122330801) with 1000 phase bins which was used to determine the phase of the main peak. The phase of the main peak at radio wavelengths is shown as a solid red vertical line.}
\label{phase0}
\end{center}
\end{figure}

The phase of the main X-ray peak was determined using a pulse profile with 1000 phase bins which was then fitted with an asymmetrical Moffat function. The explicit formula for the Moffat function is given in Appendix~C.  We also demonstrate how its shape varies when different parameters are modified. Fig~\ref{phase0} shows an example of how the phase of one Crab X-ray pulse profile (obs. ID: 0122330801) is slightly shifted in phase with respect to the radio phase (shown as a red line).

The following steps describe an example of the data reduction carried out on the \emph{XMM-Newton} data in order to assess the 
absolute timing precision:
\begin{enumerate}
\item [ ] Steps 1-3 are the same as described in Sect.~\ref{subsec:rel}
\item [4] fold the X-ray data on the radio period
\item [5] fit the X-ray pulse profile with a Moffat function
\item [6] determine the shift between the radio phase zero and the X-ray peak.
\end{enumerate}

\subsection{Evaluating the efficiency of automatic corrections made to event time jumps by the SAS}
In order to do proper timing analysis with the EPIC data, every event detected on board has to be assigned a 
correct photon arrival time. The transformation from readout sequences by the EPIC camera to photon arrival times 
of each photon is performed by the EPEA (European Photon Event Analyser, \citealt{kuster 99}). The absolute timing 
adjustment from On Board Time to UTC is done with the \emph{XMM-Newton} time correlation \citep{kirsch 04}.
A hardware problem in the EPEA can produce time jumps in some observations, which have to be corrected.

A time jump can affect the timing accuracy by broadening the $\chi^{2}$-distribution during the epoch folding 
search or by producing 'ghost peaks' \citep{kirsch 04}.

\begin{figure}[ht]
\begin{center}
\includegraphics [angle=90, bb=85 120 500 720,width=8.9cm,clip=,]{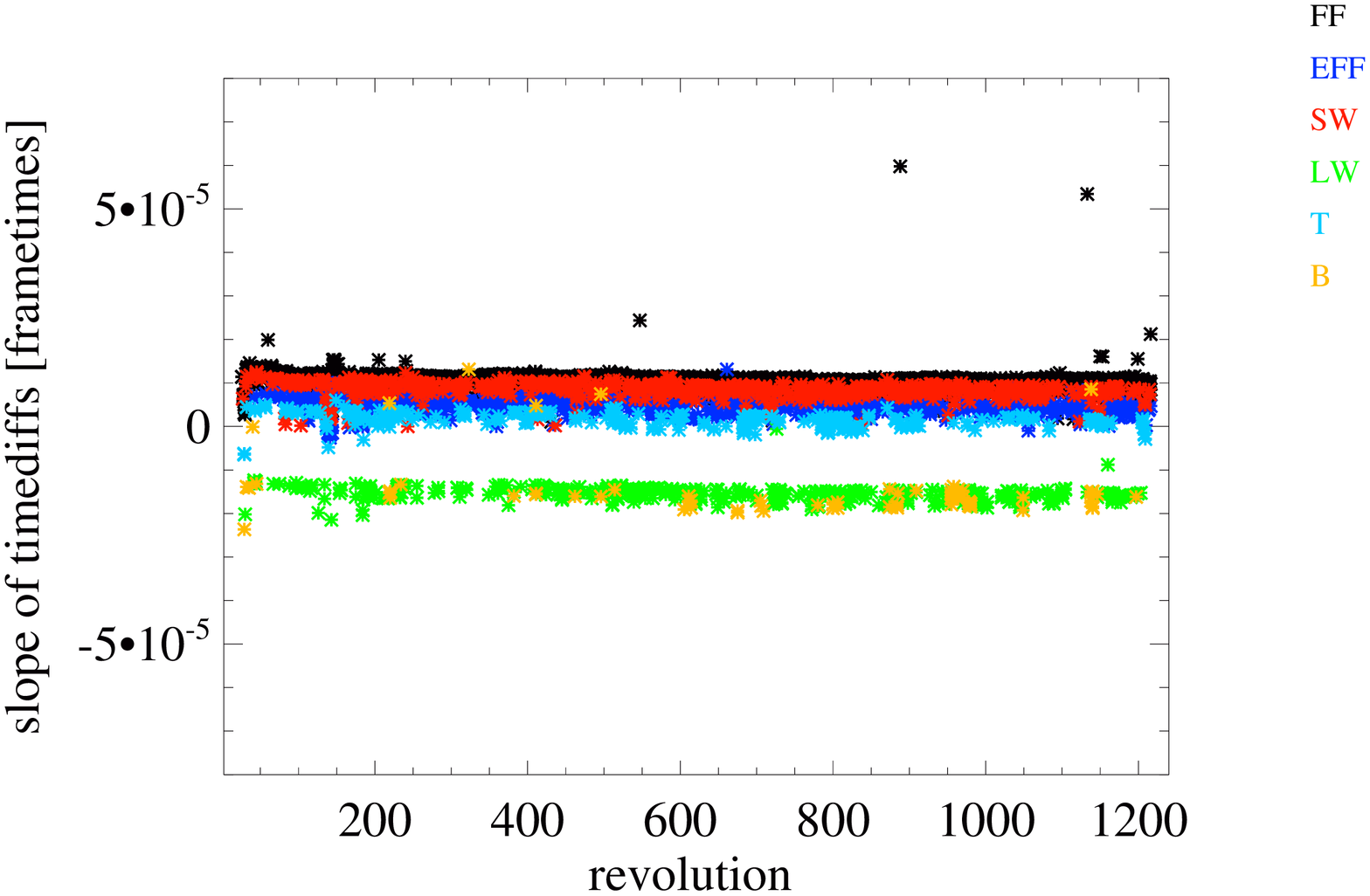}
\includegraphics [angle=90, bb=85 130 500 720,width=8.9cm,clip=,]{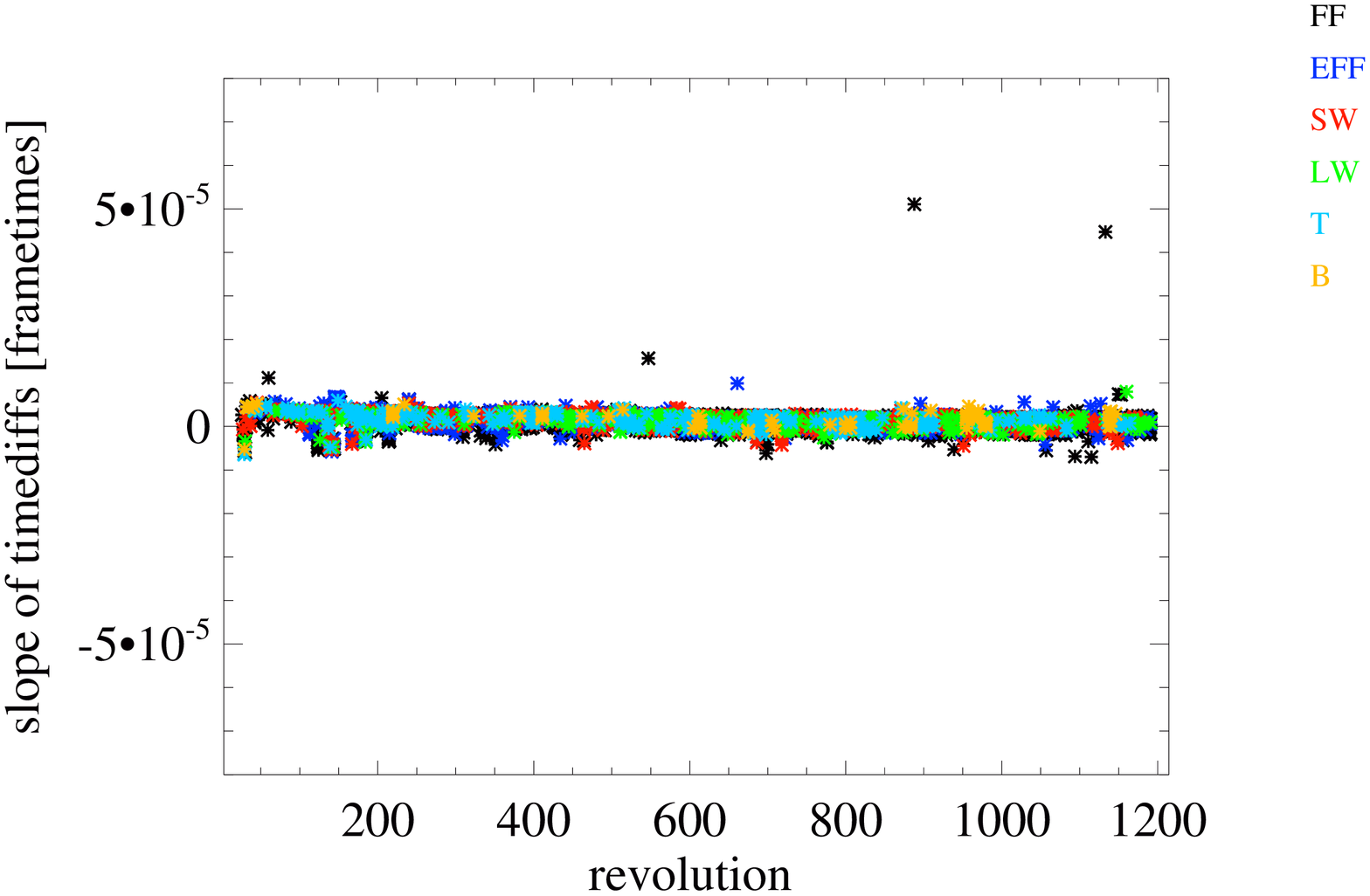}
\caption{The time difference between consecutive events determined using the slope of a linear fit to the time differences of 
consecutive events modulo the frame time. This is plotted against time.  Upper panel with old frame times, 
lower panel with the refined ones. The different modes of EPIC-pn are represented by different colours: Full Frame (black, 
Extended Full Frame (dark blue), Large Window (green), Small Window (red), Timing (light blue), Burst (yellow).}\label{slopes}
\end{center}
\end{figure}

To find time jumps reliably one needs to have as accurate CCD frame times as possible. The frame 
times for all EPIC-pn modes using all available archive data \citep{freyberg 05} were remeasured and cross-checked with an independent 
method, analysing the evolving differences between consecutive events. If these differences are correct, the slope of the relation between the difference between the arrival times of consecutive events and the same quantity modulo the frame time is zero. Fig.~\ref{slopes} shows the effect of the frame time recalibration, which brought the values of this slope very close to zero for all observational modes. Constant 'timediff' values' indicate constant frame time. 

\begin{figure} [h!]
\centering
\resizebox{\hsize}{!}{\includegraphics[angle=90, bb=40 40 551 720, clip=,]{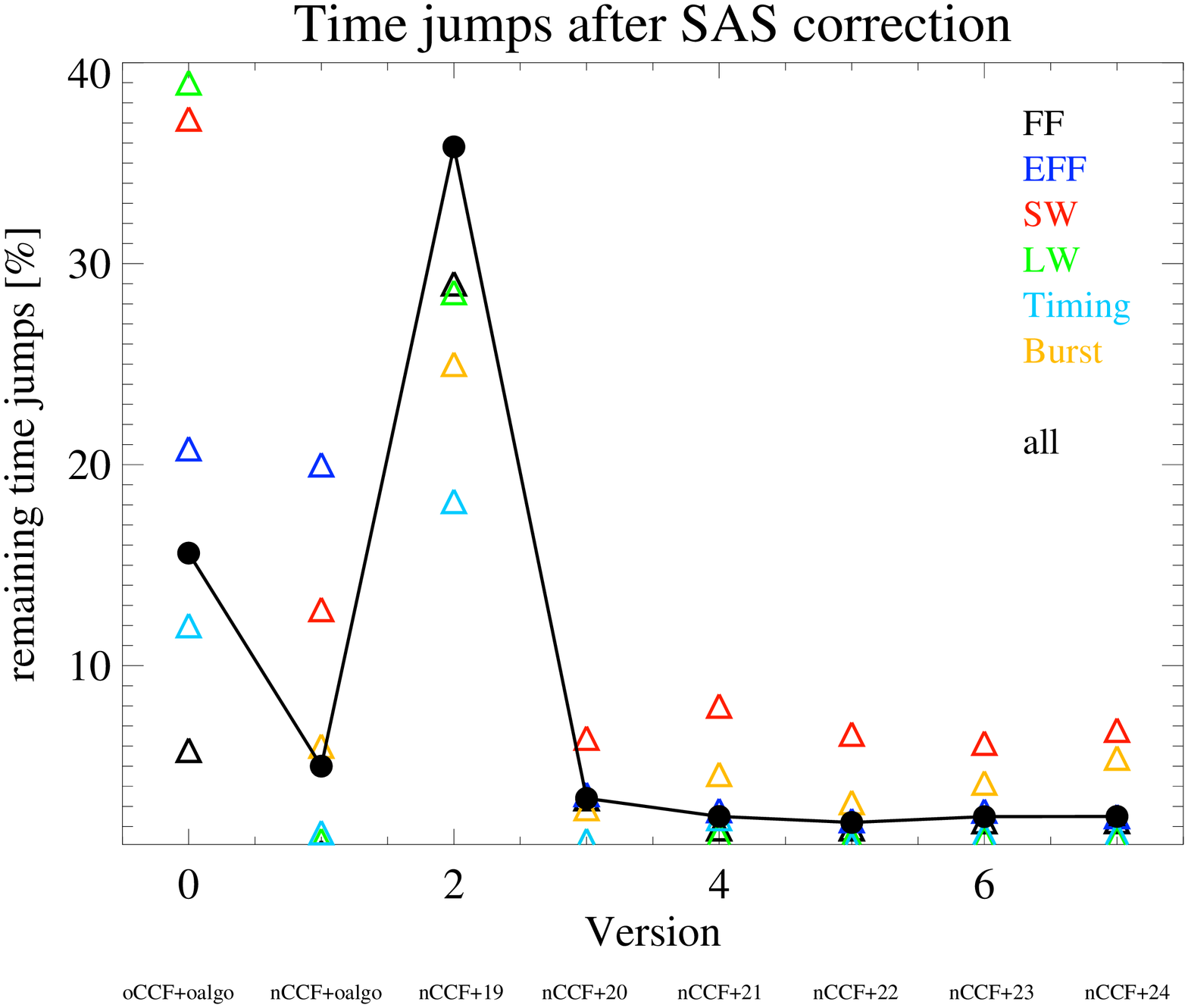}}
\caption{Remaining time jumps in the EPIC-pn data after all SAS correction algorithms. The triangles correspond to the remaining time jumps (in percent) for each Epic-pn mode: FF (black), eFF (blue), SW (red), LW green), Timing (light blue) and Burst yellow). Filled circles represent the overall remaining time jumps for all \emph{XMM-Newton} observations up to revolution 1061.
The numbers stand for the different processing versions of the algorithm.
Versions:
0: old frame times (oft) and old algorithm,   
1: new frame times (nft) and old algorithm, 
2: nft and {\tt SAS\_JUMP\_TOLERANCE} (SJT) 19.0, 
3: nft and SJT 20.0, 
4: nft and SJT 21.0, 
5: nft and SJT 22.0,
6: nft and SJT 23.0 and
7: nft and SJT 24.0}
\label{jumps}
\end{figure}

Recalibration of the time jump detection algorithm of the \emph{XMM-Newton} SAS has been done with the refined frame times. A search for time jumps in all available \emph{XMM-Newton} archive data up to revolution 1061 showed a significant number of time jumps in the data for each observational mode. The application of the new algorithm reduced the number of remaining time jumps significantly. The effects of the {\tt SAS\_JUMP\_TOLERANCE} parameter in the new algorithm (Versions 2-7) are shown in Fig.~\ref{jumps} for each EPIC-pn mode. The Version 0 shown in Fig.~\ref{jumps} represents the remaining time jumps with the old frame times and the old algorithm and it is included for comparison. Version 1 was obtained using the new frame times but with the old algorithm.
While the rate of non-corrected time jumps (averaged over all pn-modes) was 2.8 per 100 ks before 
the implementation of the SASv8.0 time jump correction, just 0.3 time jumps per 100 ks remained uncorrected after its implementation. A breakdown of time jumps for each EPIC-pn mode is given in Table~\ref{tab:tjumps}. This new improved algorithm has been implemented in the SAS as the default setting since version 8.0 ({\tt SAS\_JUMP\_TOLERANCE} = 22.0).
Table~\ref{tab:periods} indicates for which Crab observation a time jump has been corrected, and where data
had to be excluded from the analysis.

In order to identify possible remaining time jumps, the data can be processed
without the "fine-time"-correction, i.e., {\tt epframes set="infile\_pn" eventset=events.dat gtiset=tmp\_g.dat withfinetime=N}.
Then the time  $\Delta$t between successive events is calculated and divided by the frame time, \emph{FT}, of the relevant mode
(FF Mode: 73.36496 ms, 
eFF Mode: 199.19408 ms, 
LW mode: 47.66384 ms, 
SW Mode: 5.67184 ms, 
Timing Mode: 5.96464 ms,
Burst Mode: 4.34448 ms). 

A time jump is shown to exist when $\Delta$t/\emph{FT} is different from an integer by a quantity larger than a tolerance parameter. Only those time jumps which happen to be an integer multiple of the relevant \emph{FT} would not be found with this method. It is important to notice that the tolerance acceptable between $\Delta$t and the full frame time should not be bigger than (20/48828.125$\times$\emph{FT}). \\

\begin{table} [ht!]
\caption{Mean rate of residual uncorrected time jumps per 100 ks.}
\label{tab:tjumps}
\begin{tabular}{lccc}
\hline
\hline
       \noalign{\smallskip}
   pn instrument mode  & Pre SASv8.0  & Post SASv8.0 \\
                                    & per 100 ks & per 100 ks\\
  
            \noalign{\smallskip}
            \hline
            \noalign{\smallskip} 
           Full Frame (FF)             &   0.6   & 0.2  \\
	   Extended Full Frame (eFF)     &   2.1  & 0.2  \\
	   Small Window (SW)          &   3.7 & 0.7 \\
	   Large Window (LW)             &   3.9  & 0.1  \\
	   Timing                   &   1.2 & 0.1  \\
	   Burst                    &   5.3  & 0.3  \\
           \hline
	   \noalign{\smallskip}
	   Overall mean                    &   2.8  & 0.3  \\
	    \noalign{\smallskip}
            \hline

\end{tabular}
\end{table}

\section{Relative Timing accuracy of \emph{XMM-Newton}}
\label{sec:relt}
The relative timing accuracy of the \emph{EPIC-pn} camera has been studied using all six pulsars (see Fig.~\ref{alllight}), presented in Sect.~\ref{sec:targets}.
\par
As described in Appendix~B, the FWHM of the $\chi^{2}$ curve obtained during the period search analysis can be expressed in terms of the period and the exposure time of the observation (Eq.~\ref{eq:fwhm2}). From this expression, and using the Independent Fourier Space \citep[IFS;][]{emma03}, approach discussed in the appendix, an empirical formula for the error on the X-ray period was found (Eq.~\ref{eq:erperiod}).

\begin{equation}
\label{eq:erperiod}
\centering \delta P=\dfrac{\texttt{FWHM}}{dof}
\end{equation}
where \textit{dof} is the number of degrees of freedom (number of phase bins used to construct the pulse profiles minus the number of variables). The number of bins used in the pulse profiles are shown in Table~\ref{tab:epfolding}.

The relative timing accuracy was defined by Eq.~\ref{eq:rel} and therefore its error will depend mostly on how accurately the radio and X-ray periods can be measured. Other factors that could affect the relative timing accuracy are discussed in Appendix~B. Considering that the radio period measurements are more accurate than the X-ray periods (usually by 1-3 orders of magnitude; however, DM variations can sometimes cause problems and the time resolution of the radio telescopes has to be monitored), it was assumed in our analysis that their errors were negligible compared to the error on the X-ray period. Thus, it can be found that the relative timing, $\Delta$P depends exclusively on the error of the X-ray period, $\delta$P as shown in Eq.~\ref{eq:approx}.

\begin{equation}
\label{eq:approx}
\centering \Delta P \approx \delta P
\end{equation}

\begin{figure} [h!]
\centering
\resizebox{\hsize}{!}{\includegraphics[angle=90, bb=40 100 510 720, clip=,]{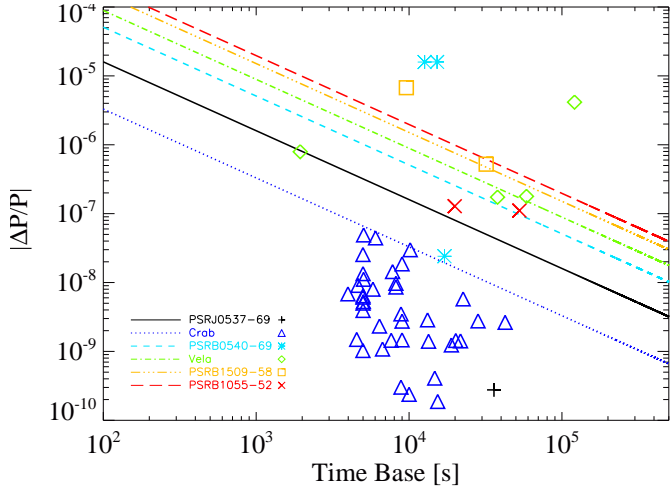}}
\caption{The relative timing accuracy of \emph{XMM-Newton's} \emph{EPIC-pn} camera.
Comparison of the expected 
$|\Delta$P/P$|$ using the assumption that the relative timing depends exclusively on the error of the X-ray period.  $\delta P$ can be empirically expressed as shown in Eq.~\ref{eq:erperiod} and \ref{eq:approx}
(given as the lines in the Figure) and the measured one obtained by comparing 
our X-ray period with the extra-(inter-)polated radio period (symbols).}
\label{rel_t_T}
\end{figure}

\begin{figure}[h!]
\centering
\resizebox{\hsize}{!}{\includegraphics[angle=90, bb=40 100 510 720, clip=,]{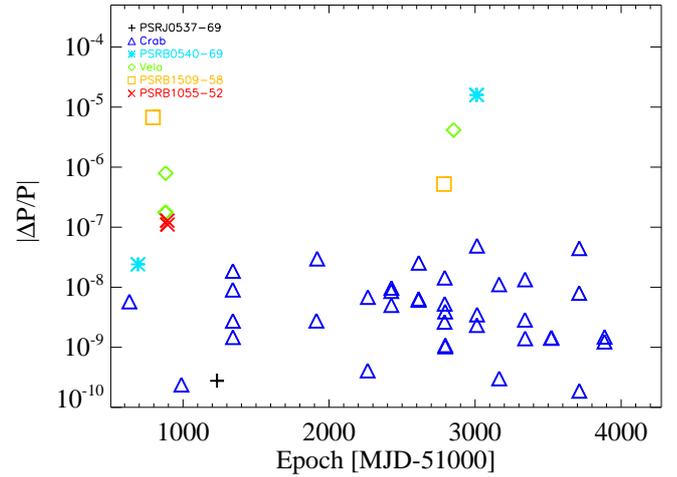}}
\caption{The relative timing accuracy of \emph{XMM-Newton's} EPIC-pn camera: 
$|\Delta$P/P$|$ for all pulsars as a function of date [MJD].}\label{arel_t_E}
\end{figure}

The relationship described in Eq.~\ref{eq:approx} allows a goodness of fit study of our measured $\Delta$P compared to the empirical $\delta$P described in Eq.~\ref{eq:erperiod}. The empirical $\delta$P was considered as an upper limit on an accurate $\Delta$P measurement. A comparison between the observed relative timing accuracy in absolute value and normalised by the corresponding period, $|\Delta$P/P$|$ (symbols) and its ``expected'' value obtained from Eq.~\ref{eq:erperiod} and Eq.~\ref{eq:approx} (lines) is presented in Fig.~\ref{rel_t_T} as a function of the exposure time and in Fig.~\ref{arel_t_E} as a function of date.

For observations where we have very reliable radio/X-ray ephemerides, the observed data points are below the 
lines of the estimated accuracies. The outliers above the respective theoretical lines for each individual pulsar 
are described below:

\begin{enumerate}

\item Radio ephemerides extrapolated over long time intervals appear to be unreliable.  Therefore, we exclude the following observations in the final calculation regarding the relative timing accuracy: Vela pulsar: 015395140; PSR~B1509-58: 0128120401; PSR~B0540-69: 0413180201, 0413180301.

\item The $\delta$P  approximation given in Eq.~\ref{eq:approx} was found to be unreliable in some cases  (e.g. for PSR~B1509-58, observation: 0312590101 and the Vela pulsar, observations: 0111080101, 0111080201). As the Vela pulsar is quite active, we may have under-estimated the error by simply extrapolating the ephemeris, however, the same can not be said about PSR~B1509-58 which is one of the most stable young pulsars known.

\end{enumerate}

As seen in Fig.~\ref{arel_t_E} there is no obvious change in the relative timing accuracy 
of the EPIC-pn camera over its lifetime.

\begin{figure} [h!]
\centering
\resizebox{\hsize}{!}{\includegraphics[angle=90, bb=50 100 515 735, clip=,]{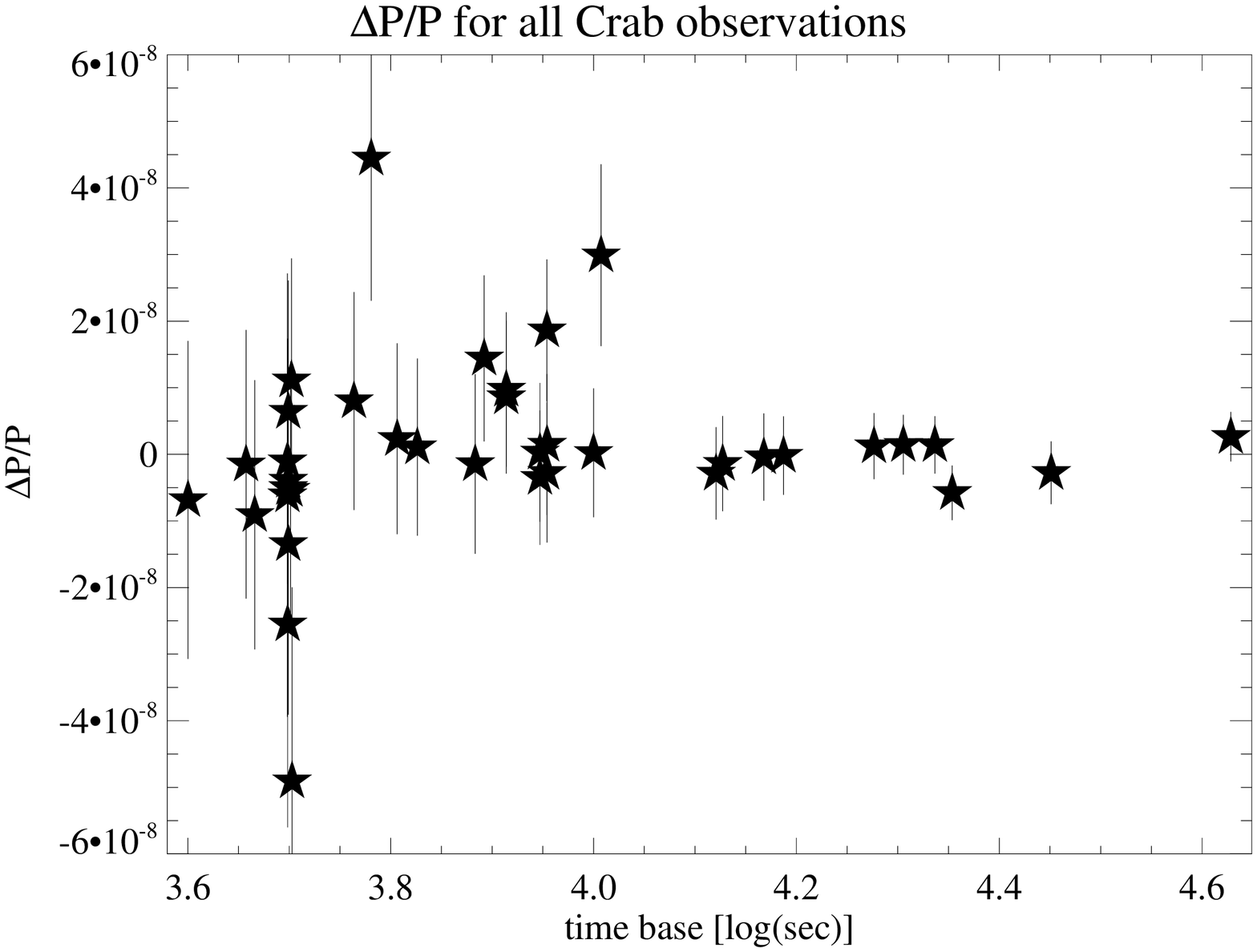}}
\resizebox{\hsize}{!}{\includegraphics[angle=90, bb=50 100 550 735, clip=,]{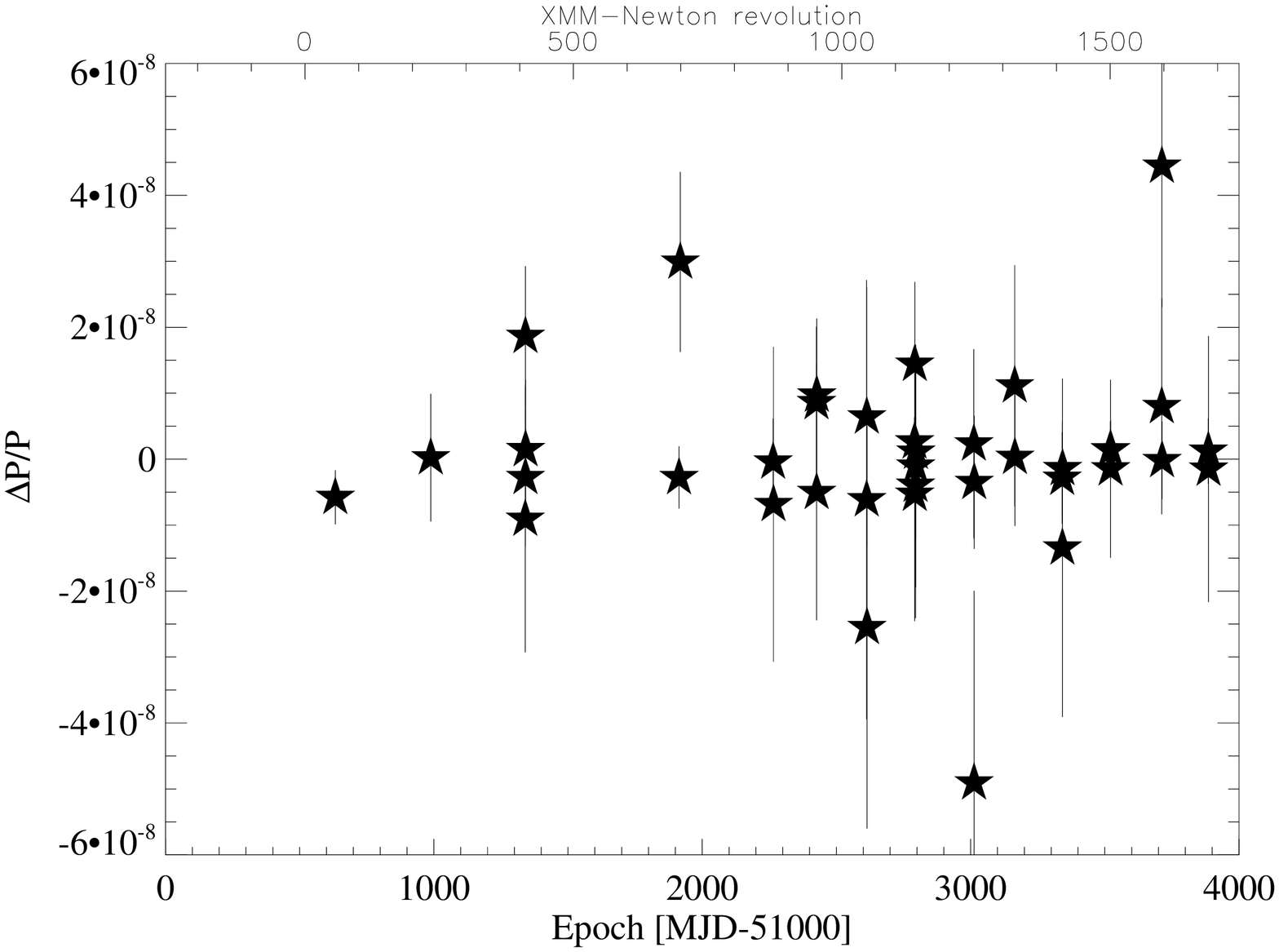}}
\caption{Relative timing using Crab monitoring data: upper panel as a 
function of observing time, lower panel as a function of MJD. These plots are regularly updated in the EPIC Calibration Status Document \citep{guainazzi} using the routine calibration observations of the Crab.}\label{crab_r_T}
\end{figure}

The results for the Crab pulsar alone are shown in Fig.~\ref{crab_r_T}. As expected, there is a tendency towards smaller uncertainties for longer observations.

For a quantitative measure of the timing accuracy the standard
deviation for the $\Delta$P/P distribution (shown in Fig.~\ref{arel_t_E}) was used.
Fitting the distributions with a Gaussian normal distribution, we found a standard deviation of $7\times 10^{-9}$ for all the pulsars together (including
the Crab pulsar) and $5\times 10^{-9}$ for the Crab pulsar alone. While
the distribution for Crab pulsar is centred at zero (within
uncertainties) the mean value of the distribution for all the pulsars combined is slightly
offset, in the sense that the X-ray period is slightly shorter on
average than the radio period. Thus, the relative deviation of the observed pulse period with respect to the most accurate 
radio data  available is $\Delta$P/P $\lesssim 10^{-8}$.

\section{Absolute timing accuracy of \emph{XMM-Newton}}
\label{sec:abst}

\begin{figure} [h!]
\centering
\resizebox{\hsize}{!}{\includegraphics[angle=90, bb=0 60 580 730,clip=]{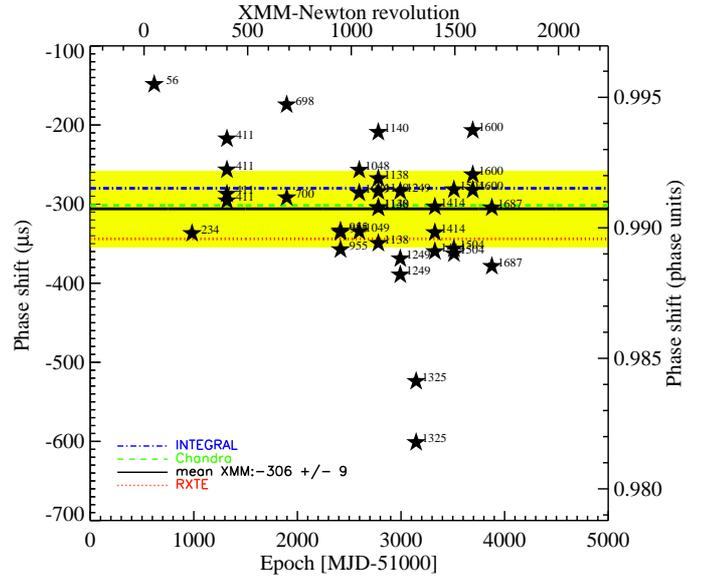}}
\caption{Absolute timing as a function of MJD using Crab-pulsar monitoring data. 
Shown are the offsets of the main X-ray peak with respect to the main radio peak in the Crab
pulse profile in time units (left scale) and phase units (right scale). The colored lines give the mean values for \emph{XMM-Newton} (solid black line, 
this work), Chandra (green dashed line), RXTE (red dot line) and INTEGRAL (blue dashed-dot line) respectively, all taken from the literature. The yellow area indicates 
the standard deviation of the \emph{XMM-Newton} data points. The superscript numbers near to each \emph{XMM-Newton} 
data point give the \emph{XMM-Newton} revolution in which the observation was carried out.}
\label{lc}
\end{figure}

The Crab pulsar shows a shift of -306\,$\mu$s (shown in Fig.~\ref{lc}) between the peak of the first X-ray pulse with respect to the radio peak.  We hereby confirm the similar results of other missions like \emph{Chandra} \citep{tennant01}, \emph{INTEGRAL} \citep{kuiper 2003}, 
and \emph{RXTE} \citep{rots 2004}. The \emph{XMM-Newton} values ("stars" in Fig.~\ref{lc}) show a considerable 
scatter with a standard deviation of 48\,$\mu$s. The formal error on the mean value of -306\,$\mu$s is $\pm\,9$\,$\mu$s.

The scatter found is consistent with the previously determined maximum integrated error for the time correlation of less than 100\,$\mu$s \citep{kirsch05}. The original requirement for an absolute timing accuracy of 1\,ms for {\em XMM-Newton}, defined before launch, has clearly been reached and even improved on by at least a factor of 20.

This scatter is likely to be due to uncertainties in the time correlation process since the phase of the main peak can be measured with an accuracy of $\mu$s. Upper limits for these processes were reported by \citet{kirsch 04}\footnote{http://xmm2.esac.esa.int/docs/documents/CAL-TN-0045-1-0.pdf} who gave a detailed description of all kinds of instrumental delays considered while converting between observing time and UTC time and estimated the spacecraft clock error to be $\sim$\,11\,$\mu$s, the uncertainty in ground station delays to be $\sim$\,5\,$\mu$s, the interpolation errors to be $\sim$\,10\,$\mu$s, the error between latching observing time and the start of frame transmission as $\sim$\,9\,$\mu$s, and the uncertainties in the spacecraft orbit ephemeris to be $\sim$\,30\,$\mu$s.  All these errors will be random for our data, and hence the fluctuations observed. The 48\,$\mu$s 1\,$\sigma$ scatter measured with respect to the mean may then be attributed uniquely to the above errors and no other systematic effect. This value can then be considered to be the minimal significant time separation between two arrival times to be considered independent.

From the initial 38 Crab observations, 32 were considered for the absolute analysis.  Six of the Crab observations were excluded for the reasons given below:

\begin{itemize}
\item Observation 0122330801: was early in the mission (rev. 56) and appears to have problems in the time correlation that can no longer be recovered.
\item Observation 0160960201: too much data had to be excluded due to time jumps which caused a dramatic reduction in the number of counts (rev 698).
\item Observations 0160960401 and 0160960601: these correspond to rev. 874 which shows the X-ray pulse peak displaced from the expected radio position.  This is likely to have been caused by a glitch shortly before the \emph{XMM-Newton} observations. This offset is more dramatic in the second observation, which has poorer statistics due to a shorter exposure time. None of these observations fall into the range shown in Fig.~\ref{lc}.
\item Observations 0412590601 and 0412590701: these correspond to rev. 1325.  The reason for the offsets is unclear. They may be due to a small, non-reported glitch or an anomalously large ground segment error, but because of the uncertainty, we excluded these observations when determining the absolute timing precision.
\end{itemize}

It was found that some observations presented pulse profiles with an excess of counts in the interpulse region of 
the Crab profile. Numerical simulations have been used to study the effect that this excess could have caused in determining the peak of the X-ray profile and thus, in determining the difference in phase between the X-ray and the radio. Using the typical 0.2-15.0 keV Crab profile as the input, 10000 light curves were 
created using Monte Carlo simulations. The strength of both peaks in the pulse profile (keeping the ratio between them constant) as well as the strength of the interpulse were selected as input parameters. 
As shown in Fig.~\ref{moffat} in Appendix~C, the Moffat fit, used to determine the phase of the main peak, 
can be used to reliably fit different tails. To fit the excess in the interpulse we 
added a Lorentzian function to the pulse profile in order to take into account the excess in the tail of the main 
peak. The secondary peak was omitted in the fit. No phase shift was found in any of the models tried, which implies that the strength of the interpulse region plays no role in determining the phase of the peak of the profile and thus all 32 of the retained Crab observations could be used to derive the absolute timing accuracy reliably.

\section{Discussion and Conclusion}
\label{sec:concl}
The Crab pulsar has been used by many missions as a calibration source for timing accuracy (\citet{kuiper 2003}, 
\citet{rots 2004}, \citet{oosterbrock}, \citet{abdo}, \citealt{molkov}). The \emph{XMM-Newton} observatory began observing the Crab pulsar during its 
 earliest orbits, monitoring its X-ray pulsation with high time resolution. 38 Crab observations spread over 10 years have been analysed in this 
 paper (from revolution 56 until revolution 1687). Measurements of the period were made with an accuracy of 
 $\sim$10$^{-11}$ s. A relative timing accuracy smaller than 10$^{-8}$ and stable with time was established for the EPIC-pn camera. This result was achieved by comparing our X-ray measurements of the Crab pulsar with high precision radio measurements at each corresponding epoch. Five isolated pulsars showing a wide range of periods and completely different pulse profiles (PSR~J0537-69, PSR~B0540-69, Vela pulsar, PSR~B1509-58 and PSR~B1055-52) were analysed to complement the study of the relative timing accuracy, confirming the results obtained with the Crab pulsar. 

For the case of PSR~B0540-69 a long term 
phase-coherent study of its period was reported by \citet{livingstone 2005b}. 
Due to its stability we considered it a good candidate to use for an
extrapolation over a long time period. As shown in Table~\ref{tab:periodso} the long extrapolation made in two of the three observations 
of PSR B0540-69 show poor results suggesting that a small glitch between the ephemeris and our observation may have 
occurred, rendering this pulsar less stable than anticipated.

An improved algorithm to detect and correct sporadic "jumps'' in the flow of the photon arrival times has been implemented with SASv8.0 \citep{guainazzi}\footnote{http://xmm2.esac.esa.int/docs/documents/CAL-TN-0018.pdf}. This method is based on a more accurate determination of the frame times for all pn modes and on a correction of frame time drifts due to temperature variation and aging of the on-board clock \citep{freyberg 05}. 
The total reduction of the rate per 100 ks of observation affected by residual uncorrected time jumps for all pn instrumental modes dropped from 2.8 before the improved algorithm to 0.3 once it was implemented.

For the absolute timing analysis, only Crab pulsar observations have been analysed since a high number of stable observations need to be considered to provide a reliable result. We have considered the phase of the first (main) peak of the X-ray profile and measured the phase difference with respect to the corresponding peak of the radio profile. Considering 32 of 38 Crab EPIC-pn observations (0.2-12 keV energy range) analysed in this paper, we confirmed previous results demonstrating that the first X-ray peak from the Crab pulsar leads the radio peak by 306\,$\pm$\,9\,$\mu$s (statistical error) with $\pm$48\,$\mu$s (1\,$\sigma$) scatter. This error is similar to the Ground Segment accuracy and defines the absolute timing accuracy of the instrument. The observed shift is consistent within 1$\sigma$ with those presented by \citet{kuiper 2003} using INTEGRAL and by \citet{rots 2004} using RXTE, as shown in Fig.~\ref{lc}.

\begin{figure} [h!]
\centering
\resizebox{\hsize}{!}{\includegraphics[angle=90, bb=60 84 530 730, clip=,]{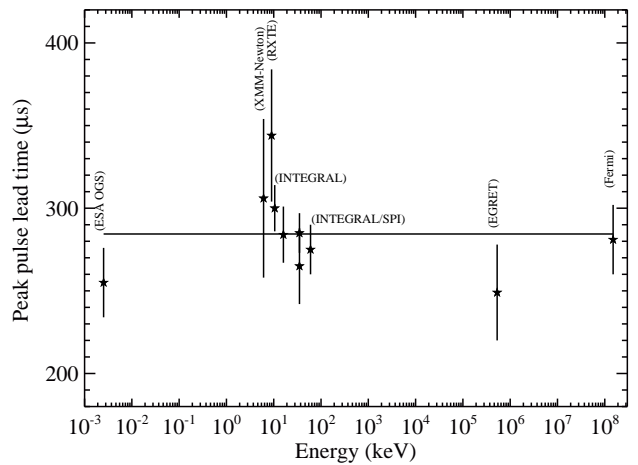}}
\caption{The peak pulse lead time ($\mu$s) of various observations are plotted against energy (keV) in optical, X-rays and $\gamma$-ray energy bands and a constant model is fitted to the data, which is found to be 284.4\,$\mu$s. The data points shown above cover over 7 orders of magnitude in energy  and come from different observations and experiments mentioned on the graph (\citet{kuiper 2003}, 
\citet{rots 2004}, \citet{oosterbrock}, \citet{abdo}, \citet{molkov}). Only the central energy corresponding to an individual observation was plotted along with corresponding lead time measured within the observed energy band.}
\label{allabsolute}
\end{figure} 

A systematic comparison of our measurements in the X-ray band with respect to other accurate measurements carried out in different energy bands from earlier observations in the optical, X-ray, and $\gamma$-ray parts of the spectrum are shown in Fig.~\ref{allabsolute}. Differences in the shifts observed  over 7 decades in energy  are marginal with an average value of 284.4\,$\mu$s. It is important to note that the large error bars quoted in the X-ray band for \emph{XMM-Newton} and \emph{RXTE} include systematic errors from the radio measurements, carried out at the Jodrell Bank Observatory.

 The origin of the electromagnetic radiation emitted from pulsars
  is still unclear.  Several models have been proposed to explain the
  origin of the high energy radiation based on different regions of
  acceleration in the pulsar magnetosphere, such as the polar cap, the
  slot gap and the outer gap models
  \citep[e.g.][]{harding78,arons79,cheng86,zhang00,harding11}.  The radio
  emission model is an empirical one and the radiation is usually
  assumed to come from a core beam centered on the magnetic axis and
  one or more hollow cones surrounding the core
  \citep[e.g.][]{rankin83}. The estimated average delay between the
  emission from differing wavelengths is therefore very significant.
  It implies that the site of radio production is distinctly different
  from that of the non-radio emission. The difference in phase between
  the radio and the X-ray radiation is about 0.008, or three degrees
  in phase angle.  This time delay of about 300\,$\mu$s most naturally
  implies that emission regions differ in position by about 90 km
  between radio and X-rays energy bands in a simplistic geometrical
  model neglecting any relativistic effects, with the radio being
  emitted from closer to the surface of the neutron star. Such high
  time resolution, high precision absolute timing, multiwavelength
  observations are therefore essential for understanding the origin of
  the pulsar emission.

\begin{acknowledgements}

  The \emph{XMM-Newton} project is an ESA Science Mission with
  instruments and contributions directly funded by ESA Member States
  and the USA (NASA). The German contribution of the
  \emph{XMM-Newton} project is supported by the Bundesministerium
  f\"{u}r Bildung und Forschung/Deutsches Zentrum f\"{u}r Luft- und
  Raumfahrt. The UK involvement is funded by the Particle Physics and 
  Astronomy Research Council (PPARC).
The Parkes radio telescope is part of the Australia
Telescope which is funded by the Commonwealth of Australia for
operation as a National Facility managed by CSIRO.

We wish to thank Dr. L. Kuiper for all his help on the absolute timing analysis procedure and results and all the discussions and comments that he provided to the development of this paper. We would also like to thank the anonymous referee for his/her useful comments and suggestions.

A. Martin-Carrillo also wish to thank the \emph{ESAC} Faculty group for their financial support during the investigation and creation of this publication.

\end{acknowledgements}

\clearpage
\begin{appendix}
\label{app:A}
\section{\emph{XMM-Newton} and its fast EPIC-pn timing modes}
\label{sec:xmm}

EPIC is capable of providing moderate energy resolution spectroscopy in the 
energy band from 0.2 to 15 keV for as many as several hundred sources in its $30'$ field-of-view.  The EPIC cameras can be operated in different observational modes related to 
different readout procedures.  Detailed descriptions of the various readout modes of EPIC-pn and their limitations are 
given by \citet{kendziorra 99}, \citet{kuster 99} and \citet{ness10}\footnote{\begin{tiny}\texttt{http://xmm.esac.esa.int/external/xmm\_user\_support/
\\
documentation/uhb/XMM\_UHB.pdf}\end{tiny}}.  

The EPIC-pn camera provides the highest time resolution in its fast Timing and Burst modes (Timing mode: 
29.52\,$\mu$s, Burst mode: 7\,$\mu$s) and moderate energy resolution ($E/\text{d}E = 10$--$50$) in the 0.2--15\,keV 
energy band. The pile-up limit (see Sec. 3.3.9 of \citet{ness10}) for a point source is 
$800\,\text{counts}\,\text{s}^{-1}$ (85\,mCrab) in Timing mode and  
$60000\,\text{counts}\,\text{s}^{-1}$ (6.3\,Crab) in Burst mode. Thus, the observations of the Crab suffer from pile-up 
only in Timing Mode, such that spectral analysis of the Crab can only be carried out accurately in Burst mode. 
However for timing purposes the effect of pile up can be neglected in the Timing mode.

\begin{figure} [ht!]
\centering
\resizebox{\hsize}{!}{\includegraphics[angle=-180, bb=80 255 530 600, clip=,]{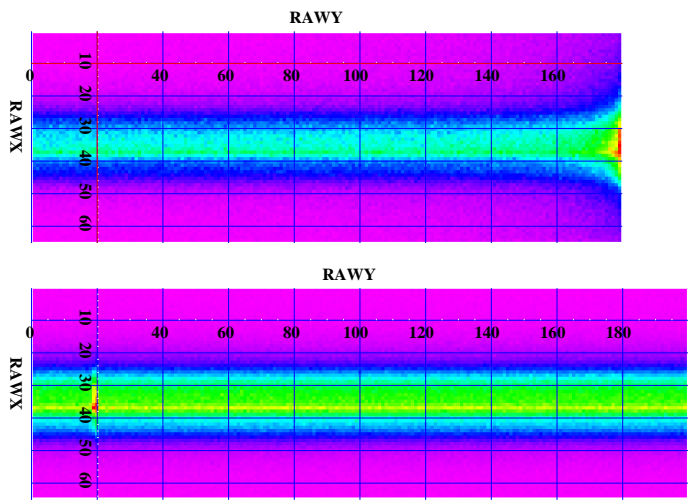}}
\caption{Upper: \emph{RAWX} and \emph{RAWY} image of a Crab observation in Burst mode. The rows 180-200 are not transmitted.
Bottom: \emph{RAWX} and \emph{RAWY} image of a Crab observation in Timing mode.
} \label{fastimages}
\end{figure}
As shown in Fig.~\ref{fastimages}, in both timing and burst modes, EPIC-pn loses spatial resolution in the shift-direction. In Timing mode,  
this is because 10 lines of events are fast-shifted towards the anodes and then the integrated signal is read out as
one line. In Burst Mode, it is because 200 lines are fast-shifted within 14.4\,$\mu$s while still accumulating information 
from the source. The stored information is then read out as normal, where the last 20 lines have to be deleted due to 
contamination by the source during the readout. The CCD is then erased by a fast shift of 200 lines, and immediately 
after that the next Burst readout cycle starts. Moreover, the lifetime in Burst mode is only 3\% and therefore, the use of this mode has been limited to
observations of very bright sources such as the Crab or X-ray transients. For our analysis we use mainly Timing 
and Burst mode observations. The images seen in Fig.~\ref{fastimages} for the Timing and Burst mode are produced 
in CCD coordinates using \emph{RAWX} and \emph{RAWY}, which are simply the pixel co-ordinates, where each pn pixel is 4.1x4.1" aside. The source appears as a stripe in the CCD RAWY direction. Source extraction regions in both modes will 
therefore always be boxes.

Operated in Timing mode EPIC-pn data show a bright line in the RAWX direction at $RAWY=19$, that is related to a 
feature in the on-board clock sequence. In the clocking scheme of the Timing mode 10 lines are shifted to the CAMEXs 
(CMOS Amplierand Multiplexer Chip) and then read out as one so called macro line,  such that the integration time 
for a normal macro line is 29.52\,$\mu$s. Within a frame time, 200 macro lines are read (corresponding to 2000 physical 
CCD lines). During the first CCD readout the first macro line contains only one CCD line and is set to bad. 
However, the integration time during the readout of the second macro-line is $29.52+23$\,$\mu$s due to electronic 
implementation of the sequencer. Therefore the integration time for a point source at $RAWY=189$ is a factor 1.8 higher 
than for all other macro line and macro line 19 receives a factor 1.8 higher flux from the point source. The feature only 
shows up for bright point sources.

There is no effect on the scientific quality of the data as long as the integration time for spectra and light curves is higher 
than the frame time in Timing mode (5.96464\,ms; \citep{freyberg 05}\footnote{\texttt{http://xmm2.esac.esa.int/docs/documents/CAL-TN-0081.pdf}}. 
Caution should be used for pulse phase spectroscopy with bin sizes below the frame time (5.96464\,ms), but only if the pulse period is a multiple of the frame time.
\end{appendix}

\clearpage
\begin{appendix}
\label{app:B}
\section{Treatment of uncertainties and reliability of radio extrapolations}
 The $\chi^{2}$ distribution obtained from the period search can be approximated by a triangle where the maximum corresponds to the true period P$_{0}$ and the points P$_{1}$ 
and P$_{2}$ where the legs of the triangle meet the level of constant $\chi^{2}$ defining the total width of the $\chi^{2}$ 
distribution. For a pulse profile with a small single peak, P$_{1}$ and P$_{2}$ can be calculated using Eq.~\ref{eq:papprox}, where 
T$_{obs}$ is the elapsed observational time and N$_{per}$ is the number of pulse periods in this time.
\begin{equation}
\label{eq:papprox}
\centering P_{1}=\dfrac{T_{\texttt{obs}}}{N_{\texttt{per}}+1}; \;\; P_{2}=\dfrac{T_{\texttt{obs}}}{N_{\texttt{per}}-1}
\;\;\; \texttt{where} \; N_{\texttt{per}}=\dfrac{T_{\texttt{obs}}}{P}
\end{equation}
\par
For a triangular function the Full Width Half Maximum (FWHM) is equal to $(P_{2} - P_{1})/2$ and can be expressed 
as in Eq.~\ref{eq:fwhm2} as a function of the period and the elapsed observation time.
\begin{equation}
\label{eq:fwhm2}
\centering \texttt{FWHM}=\dfrac{P_{2}-P_{1}}{2} \Rightarrow
\texttt{FWHM}=\dfrac{P^{2}}{T_{\texttt{obs}}}
\end{equation}
\par

\begin{figure} [ht!]
\centering
\resizebox{\hsize}{!}{\includegraphics[angle=90, bb=40 115 510 700, clip=,]{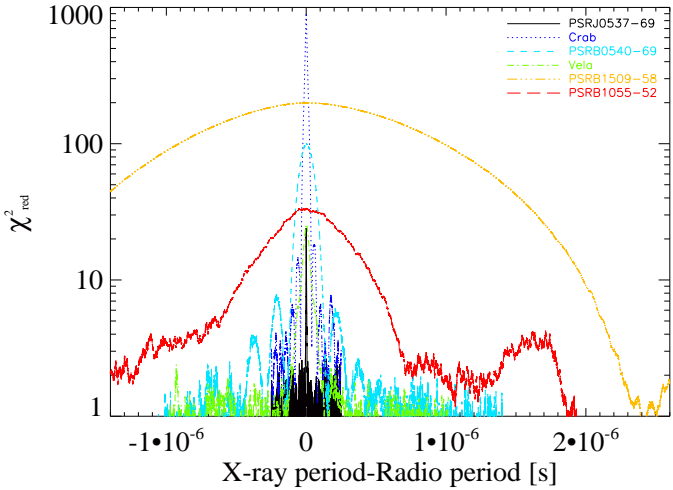}}
\caption{Sample $\chi^{2}$ distributions for one observation of each of the 
studied pulsars. } \label{allchi2}
\end{figure}

\begin{figure} [ht!]
\centering
\resizebox{\hsize}{!}{\includegraphics[angle=90, bb=40 115 510 720, clip=,]{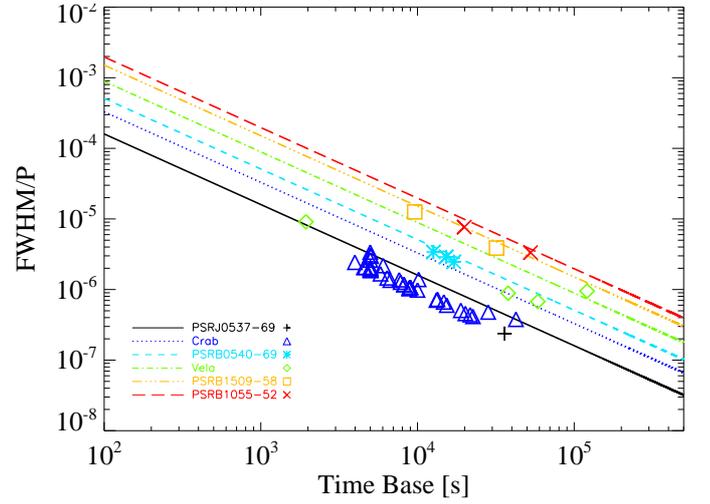}}
\caption{Comparison of the predicted FWHM of the $\chi^{2}$ 
distributions (lines) and the observed ones (symbols). All values were normalised using the pulsar period to be able to present all the pulsars on the same diagram.} \label{fwhm_comp}
\end{figure}

An expected FWHM of the $\chi^{2}$ distribution can be estimated using Eq.~\ref{eq:fwhm2}. A comparison of such estimations (lines) and the measured FWHMs from all the observations is shown in Fig.~\ref{fwhm_comp}. All values were normalised using the pulsar period to be able to present all the pulsars on the same diagram. All measured values of FWHM/P are about a factor of 3 smaller than those predicted  for the Crab and Vela pulsars. In the other four pulsars the ratio between the measured and predicted values is $\sim$\,1.3. This would suggest that this approximation works better in single peaked pulse profiles.

Empirically, two periods can be considered completely independent
from each other when their difference is at least P$^{2}$/T (one Independent Fourier Space, IFS \citep{emma03}), which is identical to
the FWHM definition in Eq.~\ref{eq:fwhm2}. One IFS can be seen, then, as the delta in pulse period which will smear a perfect pulse profile to a flat profile when folded over the complete observing time $T_{\texttt{obs}}$.
This approach is quite conservative and smaller changes than one IFS in period can be easily seen. We have 
found that a rough estimate of the uncertainty in the measured period can be found by dividing the FWHM by the number of phase bins used to construct the pulse profile (degrees of freedom, see Table~\ref{tab:epfolding}). Thus, two periods will be considered different when the pulse profile is smeared by one bin instead of one whole phase. The error on the X-ray period can then be written as shown in Eq.~\ref{eq:aperper}.

\begin{equation}
\label{eq:aperper}
\centering \delta P=\dfrac{\texttt{FWHM}}{dof}
\end{equation}

Besides providing a good estimate of the error on the period, the Independent Fourier Space approach can also provide a good indication of how reliable the extrapolation (or interpolation in the case of the Crab pulsar) of a period can be. Since we defined the relative timing accuracy in Sect.~\ref{subsec:rel} based on the reference period (normally obtained from radio observations) at the time of the \emph{XMM-Newton} observation, it is critical to understand how reliable, and in some form, how accurate, this parameter really is. For clarification, and due to the huge amount of data available we will focus on the Crab pulsar only. However the same applies to all the pulsars studied in this paper.

Eq.~\ref{eq:aperper} establishes that two periods are completely different if the pulse profile is smeared by one bin. By studying the phase smear, we can determine whether the pulse profile has been affected by a glitch or whether extrapolating the ephemeris over (long) time periods leads to inaccuracies.

If a simple period evolution with time (including the second derivative) is assumed, the phase smear is then defined as in Eq.~\ref{eq:phsmear}.

\begin{equation}
\label{eq:phsmear}
\centering \texttt{Phase Smear}=\dfrac{(P_{\texttt{extrap}}-P_{0})T_{\texttt{obs}}}{P_{\texttt{extrap}}\times P_{0}}
\end{equation}
where P$_{\texttt{extrap}}$ is the extrapolated period at the time T$_{0}$, P$_{0}$ is the actual period at that time and T$_{\texttt{obs}}$ is the exposure time of the observation.

\begin{figure} [!ht]	
\centering
\includegraphics[angle=90, width=.45\textwidth]{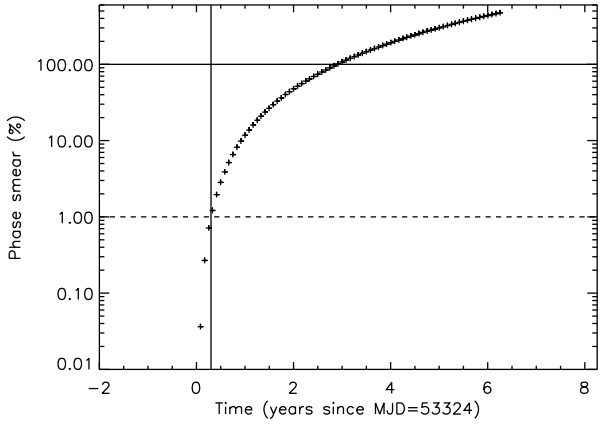}
\caption{Phase smear of the Crab pulsar versus time. The timing noise seems to be a prominent feature for the Crab pulsar and therefore extrapolation over periods of 4 months will already reach our 1\% limit (dashed line).}
\label{phaseSmear}
\end{figure}

Using radio data from the Jodrell Bank Observatory, the phase smear versus time of extrapolated periods is shown in Fig.~\ref{phaseSmear}. For the relative timing analysis 100 phase bins have been used and therefore a limit of 1\% smearing is imposed by the criteria described above. Extrapolating the period, the Crab pulsar would reach that limit in 4 months. Considering that the Jodrell Bank Observatory provides an updated ephemeris every month, the actual smearing effect will be much lower ($\sim$\,0.1\%) and other properties such as timing noise will not affect our relative timing analysis. For the Crab, we actually used interpolation rather than extrapolation, see Sect.~\ref{subsec:rel}, so the phase smear was further minimised ($\sim$\,0.09\%).

\end{appendix}

\clearpage
\begin{appendix}
\label{app:C}
\section{The Moffat function}

\begin{figure} [!ht]
\centering
\resizebox{\hsize}{!}{\includegraphics[angle=90]{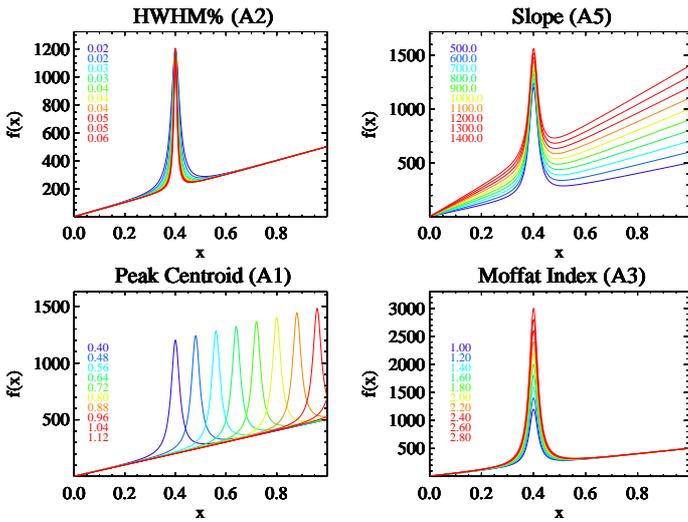}}
\caption{Variations in the shape of the Moffat function when parameter values are changed.}
\label{moffat}
\end{figure}

The Moffat function is a modified Lorentzian with a variable power law index \citep{moffat 69}. In Fig.~\ref{moffat} the 
behaviour of the function is shown as a function of its parameters. The function presents different tails on each side of the maximum which fit the main pulse of the Crab profile better than a normal Lorentzian or Gaussian function. The explicit formula of the Moffat function is the following:

\begin{equation}
\centering y={\dfrac{A_{0}}{(((x-A_{1})/A_{2})^{2}+1)^{A_{3}}}+A_{4}+A_{5}x}
\end{equation}
\begin{flushleft}
The different parameters represent: 
\end{flushleft}
\begin{itemize}
\item[ ] $A_{0}$: normalization 
\item[ ] $A_{1}$: Peak Centroid 
\item[ ] $A_{2}$: HWHM 
\item[ ] $A_{3}$: Moffat index 
\item[ ] $A_{4}$: offset 
\item[ ] $A_{5}$: slope 
\end{itemize}

The variation in the shape of the Moffat function for different values of the important parameters is shown in Fig.~\ref{moffat}. Upper left: $A_{2}$  changes from 0.02 to 0.06; upper right: $A_{5}$ changes from 500 to 1400; lower left: $A_{1}$ changes from 0.4 to 1.12; and lower right: $A_{3}$ changes from 1.0 to 2.80. \\

\end{appendix}
\end{document}